\documentclass[11pt]{article}

\usepackage[utf8]{inputenc}
\usepackage{amsmath,amssymb,amsthm}
\usepackage{graphicx}
\usepackage{subcaption}
\usepackage{bbm}
\usepackage{tikz}
\usepackage{dsfont}
\usepackage{enumerate}
\usepackage{bm}
\usepackage{algorithm}
\usepackage{algpseudocode}
\usepackage[colorlinks=true, linkcolor=blue, citecolor=blue, urlcolor=blue]{hyperref}
\usepackage[margin=1in]{geometry}

\newtheorem{theorem}    {Theorem}[section]   
\newtheorem{lemma}      [theorem]{Lemma}     
\newtheorem{assumption} [theorem]{Assumption}
\newtheorem{remark}     [theorem]{Remark}    

\newtheorem{corollary}  [theorem]{Corollary}

\DeclareMathOperator{\KL}{KL}

\def\R{{\mathcal{R}}}
\def\B{{\mathbf{B}}}

\algrenewcommand\algorithmicrequire{\textbf{Initialization:}}
\algrenewcommand\algorithmicensure{\textbf{Output:}}

\title{Accelerated Langevin Sampling with Birth-Death Process and Exploration Component}
\author{
  Lezhi Tan\thanks{Stanford University, \texttt{lezhitan@stanford.edu}} 
  \and 
  Jianfeng Lu\thanks{Duke University, \texttt{jianfeng@math.duke.edu}} 
}

\date{} 

\begin{document}

\maketitle

\begin{abstract}
Sampling a probability distribution with known likelihood is a fundamental task in computational science and engineering. Aiming at multimodality, we propose a new sampling method that takes advantage of both birth-death process and exploration component. The main idea of this method is \textit{look before you leap}. We keep two sets of samplers, one at warmer temperature and one at original temperature. The former one serves as pioneer in exploring new modes and passing useful information to the other, while the latter one samples the target distribution after receiving the information. We derive a mean-field limit and show how the exploration component accelerates the sampling process. Moreover, we prove exponential asymptotic convergence under mild assumption. Finally, we test on experiments from previous literature and compare our methodology to previous ones. 
\end{abstract}

\bigskip

\noindent\textbf{Keywords:} Markov chain Monte Carlo, birth-death, exploration process

\medskip

\noindent\textbf{MSC Codes:} 65C05, 65C35, 60J80, 62F15

\section{Introduction}
\label{sec:1}

Sampling high-dimensional distributions has widespread applications in fields such as machine learning, Bayesian inference, and molecular simulations. The most well-known approach is to construct a Markov chain with an invariant distribution that agrees with the target distribution, such as Random-Walk Metropolis \cite{gelman1997weak,roberts2001optimal}, Unadjusted Langevin Algorithm and Metropolis Adjusted Langevin Algorithm \cite{roberts1996exponential,roberts1998optimal,roberts2002langevin,Alain2019Bayesian}, Hamiltonian Monte Carlo \cite{Alexandros2013Hamilton,chen2014stochastic}, and others. These methods have been proven to be highly efficient when the log-likelihood is convex, see for example \cite{dwivedi2018log, durmus2022geometric, wu2022minimax, lee2021structured} and references therein. However, in reality, probability distributions are seldom logconcave and are often multimodal. Multimodality has long been a challenge in sampling, as it can impede samplers from jumping across low-probability regions to reach new high-probability regions. Under multimodality, most MCMC approaches based on localized proposing mechanisms fail to explore the entire state space efficiently, resulting in biased outputs \cite{gelman1997weak,roberts2001optimal}. As a result, there is a great need for the development of efficient sampling methods for multimodal distributions.\par

For more efficient sampling algorithms, one hopes for faster convergence to the target distribution $\pi$, starting from some initial distribution.  
This objective can be broken down into three subtasks.
The first is to explore the entire state space, which means finding all high-probability regions (or modes). The second is to make $\rho$, the sampling distribution, converge locally to $\pi$ inside every high-probability region. The final task is to balance the weights of $\rho$ in each high-probability region. The second task is equivalent to sampling a unimodal distribution, which is often log-smooth and log-concave. This is relatively easy, as there are many methods efficient for log-concave sampling and their convergence has been well-studied for decades \cite{dwivedi2018log,lee2021structured,cheng2018underdamped}. However, the first and the third ones are much more difficult when the target distribution exhibits multimodality.

Concerning the first task, tempering approach, in particular parallel tempering (PT) \cite{geyer1991markov}, is well-known for accelerating exploration. It involves constructing multiple parallel Markov chains that individually converge to tempered versions of the target distribution $\pi^{\beta}(x)$ for $\beta<1$.  At these warmer temperature levels, the multimodality of the target distribution is reduced, which allows the samplers to more easily jump across low-probability regions and explore the entire state space. 
On the other hand, while faster exploration at warmer levels may occur, it does not necessarily translate to faster exploration at the target level, because swapping between different temperature levels can be inefficient. While this can be overcome by increasing the number of temperature levels,  running multiple levels of Markov chains simultaneously requires more memory and longer computation time.  \par

In addition to the technique of swapping, there are other methods that leverage information from higher temperature Markov chains. One such approach is used in ALPS (Annealed Leap-Point Sampler), which was recently proposed by Tawn et al. \cite{tawn2021annealed}. It first uses optimization to locate modes at a higher temperature, and then directly integrates these modes into Metropolis-Hastings updates. This is in contrast to the traditional method of maintaining several parallel chains at descending temperatures. In our work, we adopt this approach and refer to it as the Exploration Component.\par

The third task involves adjusting the weights between different high probability regions, which requires transporting samplers from one region to another. One possible way is the birth-death process proposed in \cite{lu2019accelerating}. While the Langevin Dynamic combined with Birth-Death process (BDLS), as shown in \cite{lu2022birth}, can achieve asymptotic exponential convergence rate independent of the spectral gap, the actual (pre-asymptotic) performance is limited by the exploration stage - where new modes of the distribution are discovered by the sampler. Only after all modes are discovered, BDLS becomes efficient in adjusting the samplers across them, and hence converges to the target distribution. Thus, while BDLS resolves the third task, it hinders upon the first task of exploration.

Taking into account the strengths and limitations of existing methods, we propose a combined algorithm that addresses the three tasks of exploration, convergence inside high probability regions, and balancing weights between them. Our new sampling algorithm, named BDEC, is based on Langevin sampling and incorporates the birth-death process (BD) and the exploration component (EC). To accomplish the first task, we use samplers at a warmer temperature to explore the state space, and employ EC to detect new modes and update the collection of modes. We then insert samplers around new modes via a Metropolis-Hastings step. Once the insertion is successfully done, we return to within-temperature moves at the original temperature using Langevin MCMC with BD. In this combination, Langevin ensures local convergence within each high probability region, while BD helps achieve efficient balancing of weights, independent of the degree of multimodality.

\paragraph{Our contribution}
We propose a new sampling scheme BDEC, which combines BD and EC to accelerate Langevin dynamics. Through numerical experiments, we show that using BD and EC together dramatically accelerates the Langevin dynamics. Our algorithm has faster convergence rate than both BDLS and ALPS, and has lower computational costs than ALPS. \par
For theoretical analysis, we consider our Markov Chain under mean-field limit and continuous time limit. We prove that the asymptotic convergence is exponential, which does not depend on the spectral gap. 
In particular, our analysis not only applies to the asymptotic regime where the state space is fully explored, but also taking into consideration the exploration phase. In comparison, most previous literature directly assume that the starting distribution has already covered the entire state space.

\paragraph{Related works}
Two important components in BDEC, the birth-death process and exploration component, are directly inspired by previous literature. The birth-death process allows for mixing between Markov chains, and is generated from non-local mass transport dynamics known as birth-death dynamics. \cite{lu2019accelerating} first combined the birth-death dynamics with Langevin dynamics, and its improved analysis was given by \cite{lu2022birth}. The birth-death dynamics has also been applied to accelerate training for neural networks in mean-field limit \cite{rotskoff2019neuron}. 
There is a large body of literature on sampling schemes with interacting particles, including BDLS \cite{lu2019accelerating}, ensemble MCMC \cite{goodman2010ensemble,lindsey2022ensemble} and SVGD \cite{liu2017stein,chewi2020svgd}. Although these methods can transport particles inside the explored area, their exploration speed is limited. \par

To accelerate the exploration process, we utilize an extra Markov chain at warmer temperature. 
The use of fact that exploration is faster at warmer temperature can date back to the distinguished Parallel Tempering \cite{geyer1991markov, marinari1992simulated}. 
Recent improvements to PT include the optimal choice of temperature spacings \cite{tawn2018optimal}, accelerating mixing between temperatures \cite{tawn2019accelerating}, using a weight-preserving version of tempered distributions \cite{tawn2020weight}, and a variant with infinite swapping limit \cite{lu2019methodological}. Instead of swapping between multiple levels of Markov chains like PT, we only keep one extra Markov chain and extract useful information about modes, similar to the exploration component in ALPS \cite{tawn2021annealed}. We generate proposals according to Gaussian mixture approximations, and keep a reversible chain according to the distinguished Metropolis-Hastings mechanism \cite{10.2307/2334940}. Gaussian approximation based on mode location and Hessian has been widely used in Bayesian inference, MCMC and stochastic process \cite{opper2009variational,quiroz2022gaussian}. 
Our use of Gaussian mixture approximation shares the same idea with mode jumping MCMC \cite{tjelmeland2001mode} that we both use optimization to locate modes and then construct approximations to target distribution based on Hessian at the modes. Mode jumping MCMC also allows for non-local movements, however, its exploration process is not augmented. \par

The exploration component was first proposed as a key part of ALPS algorithm in \cite{tawn2021annealed}, followed by theoretical analysis in \cite{roberts2022skew}. However, our use of the exploration component is not identical to ALPS. While \cite{tawn2021annealed} addressed the problem of low Metropolis-Hastings acceptance rates caused by significant skewness in the target distribution, they did so by using annealing \cite{bertsimas1993simulated,neal2001annealed}. Specifically, they employed several annealed Markov chains at colder temperatures, performing Metropolis-Hastings only at the coldest temperature, and swapped samplers between different temperature levels to insert new modes into the target level. Inevitably, this strategy leads to large memory and computational costs, similar to parallel tempering (PT). In contrast, we show in our work that annealing is not necessary for our approach. The use of the Birth-Death process means that even with relatively low MH acceptance rates, a single accepted proposal can result in immediate transportation of enough samplers to new modes.

\section{The Algorithm}
\label{sec:Algorithm}

Before we present the algorithm, let us first collect the notations used. 
\subsection*{Notations}
\begin{itemize}
	\item $\pi(x) \propto e^{-V(x)}$, the target probability density; $\pi_{\beta}(x) \propto e^{-\beta V(x)}$, the tempered version that shares the same modes with $\pi$. When $\beta = 1$, $\pi_{\beta}$ corresponds to the target distribution. Here and in the sequel, we will abuse notation of distribution with its density; 
	\item $\beta_{\text{hot}} < 1$, the inverse temperature of the warmer level for exploration;
    \item $m_t$, the number of mode points discovered by time $t$; when no confusion might occur, it will be also abbreviated as $m$;
	\item \label{definitionofM}  $M^t:=\{ \mu_1 , \ldots, \mu_{m_t} \}$, the mode locations of $\pi$ discovered by iteration $t$;
 \item $S^t:=\{ \Sigma_1 , \ldots, \Sigma_{m_t} \}$, the collection of covariance matrix, 
 $$\Sigma_i= -\left[\nabla^2 \log (\pi(\mu_i))\right]^{-1},~~i=1,\ldots m_t;$$
 \item $W^t=\{\widehat{w}_1,\ldots, \widehat{w}_{m_t} \}$, the collection of weights of modes,
    \begin{equation}
    \label{eq:defweight}
        \widehat{w}_j= \frac{\pi(\mu_j) |\Sigma_j|^{1/2}}{\sum_{k=1}^{m_t} \pi(\mu_k) |\Sigma_k|^{1/2}}, ~~j=1,\ldots,m_t;
    \end{equation} 
	\item  $I^t = \{M^t,S^t,W^t\}$ the information of modes collected by iteration $t$. We also abbreviate these notations as $I = \{M,S,W\}$ when there is no confusion;
	\item $\hat{\rho}(\cdot ;I)$, the Gaussian mixture approximation of $\pi$ based on the information of modes $I$, 
	\begin{equation}
    \label{eq:defproposal}
	\hat{\rho}(x ;I) = \sum_{i=1}^m \widehat{w}_i \mathcal{N}(x ;\mu_i,\Sigma_i); \end{equation}
    \item $P(x;y, I)$, the corresponding transition probability from $y$ to $x$ in Metropolis-Hastings updates:
 $$ P(x; y,I) = \hat{\rho}(x;I) A(x,y) + \mathbf{1}\{x=y\} \Bigl(1- \int \hat{\rho}(s;I) A(s,y) ds \Bigr),
 $$
 where $A(x,y)$ denotes the acceptance rate of proposing $x$ from $y$,
 \begin{equation} \label{eq:acceptancerate}
     A(x,y) = \min \Bigl\{1, \frac{\hat{\rho}(y;I) \pi(x)}{\hat{\rho}(x;I) \pi(y)} \Bigl\};
 \end{equation}
 	 
	\item $X^t:= \{ x^t_1,x^t_2,\ldots, x^t_{N} \}$, $Y^t = \left\{y^t_1,y^t_2,...,y^t_{\widehat{N}}\right\}$, the collections of particles at original temperature and the warmer temperature by iteration $t$, respectively, where $N(\widehat{N})$ denotes the number of particles at target (tempered) level.  In most cases, we will let $N \geq \widehat{N}$;
 
	\item $K(\cdot,\cdot)$, the kernel function; $\rho_t(\cdot)$, the kernalized density of $X^t$:
	\begin{equation} \label{itemrho}
    \rho_t(\cdot) = \frac1N \sum_{i=1}^N K(x_i^t,\cdot).
	\end{equation}
\end{itemize}

\subsection{Algorithm outline}
The BDEC algorithm utilizes two groups of samplers $X,Y$ at two different temperature levels: At target level, set as $\beta = 1$, $X$ samples the target probability distribution $\pi(x)$; at warmer level,  $\beta_{hot} < 1$, $Y$ samples the tempered version of probability distribution $\pi_{\beta}(x)$. Samplers $Y$ explore the state space faster than $X$ thanks to its higher temperature. \par
There are two kinds of moves: within-temperature moves and interaction between two levels. We use birth-death accelerated Langevin sampling (BDLS) \cite{lu2019accelerating} for within-temperature updates of $X$, and unadjusted Langevin algorithm for within-temperature updates of $Y$.  

After every $T$ within-temperature moves, the exploration component collects the information provided by $Y$ and passes it to $X$, which is interpreted as interaction between two levels. First, the algorithm locates a bunch of modes by optimization on the potential surface starting at a subset of $Y$, aiming to find new modes (local minima). These modes will help the sampler $X$ by proposal of new locations incorporating such modes together with a Metropolis-Hastings acceptance-rejection to ensure that $X$ samples the correct distribution. 
Benefit by the exploration component, $X$ is not limited by the exploration given by the Langevin dynamics at the original temperature.

The outline of the BDEC procedure is given in Algorithm~\ref{alg:BDEC}, the details of its constituent parts are elaborated in the following subsections~\ref{Sec:moveofY}, \ref{Sec:exp} and \ref{Sec:ALSBD}. The exploration component is introduced in subsection~\ref{Sec:exp}, while the birth-death process is introduced in subsection~\ref{subsec:BDLS}.

\begin{algorithm}
  \caption{Langevin Sampling with Birth-Death Process and Exploration Component}
    \label{alg:BDEC}
  \begin{algorithmic}[1]
    \Require{Hotter temperature $\beta_{\text{hot}}$; initial mode point information $I^{0}:=\{  M^0,S^0,W^0\}$; initial particles at the target temperature and the warmer temperature, $X^0=\{x^0_i\}_{i=1}^N,Y^0 = \{y^0_j\}_{j=1}^{\widehat{N}}$; a Metropolized transition operator, $P(\cdot;x, I)$;
$J$ the number of total iterations, and 
$T$ the number of within-temperature moves in each iteration.} \Procedure{BDEC}{$X^0,Y^0$}

      \For{$j \gets 1 \textrm{ to } J$}
      		\State{$m \gets (j-1)T$}
			\State{\textbf{1. Within-temperature Updates at Tempered Level:}}				 \For{$t \gets 1 \textrm{ to } T$}
			
			 \State{Update: $Y^{t+m}\gets \text{Langevin}(Y^{t+m-1}; \beta = \beta_{\text{hot}})$} 
			
			\EndFor
			
			\State{\textbf{2. Exploration Component:} (apply Algorithm~\ref{alg:explore})}				\State{\begin{equation*}
				I^{j}, \text{New} \leftarrow \text{ModeFinder}(Y^{jM}, I^{j-1}) 
				\end{equation*}}
			\State{\textbf{3. Within-temperature Updates at Target Level: }}
			\If{New == True}
			\For{$t \gets 1 \textrm{ to } T$}
			\State{Update: $X^{t+m-1/2} \gets P(\cdot;X^{t+m-1}, I^{j})$}
			\State{Adjust: $X^{t+m} \gets BD(X^{t+m-1/2},\cdot)$}
			\EndFor
			\Else{ New == False}
			\State{Apply BDLS Algorithm~\ref{alg:BDLS}:}
			\For{$t \gets 1 \textrm{ to } T$}
			\State{Update: $X^{t+m-1/2} \gets \text{Langevin}(X^{t+m-1}; \beta = 1)$}
			\State{Adjust: $X^{t+m} \gets BD(X^{t+m-1/2},\cdot)$}
			\EndFor
			\EndIf

      \EndFor
      \State \Return{$X^{JT}$}
    \EndProcedure
  \end{algorithmic}
\end{algorithm}

\subsection{Within-temperature Updates at Tempered Level}
\label{Sec:moveofY}
The first stage in each iteration is the within-temperature moves of $Y$. We hope $Y$ to serve as a pioneer in exploration, which means, samplers in $Y$ should not be trapped in one high-probability region as long as those in $X$. Meanwhile, we hope samplers in $Y$ do not jump away from any new high-probability region too soon, but to stay until information of this new region is successfully passed to $X$. For these reasons, we set $Y$ as samplers of the tempered distribution $\pi_{\beta_{\text{hot}}}$. $\beta_{\text{hot}}$ is a parameter to be tuned. As $Y$ do not need to accurately  converge to $\pi_{\beta_{\text{hot}}}$, we simply choose unadjusted Langevin algorithm  \cite{roberts1996exponential} for its update (see algorithm~\ref{alg:ULA}). One could also choose other MCMC approaches in the implementations. 

 \begin{algorithm}[ht]
  \caption{Unadjusted Langevin algorithm (ULA)}
    \label{alg:ULA}
  \begin{algorithmic}[1]
	\Require{$V(x)$, the log-likelihood of target distribution $\pi(x)$; inverse temperature, $\beta$; an initial set of particles $\{y^{0}_i\}_{i=1}^{\widehat{N}}$; $T$ the number of updates, $\Delta t$ the time step.}
    
	\For{$t \gets 1 \textrm{ to } T$}
	\For{$i \gets 1 \textrm{ to } \widehat{N}$}
    \State{set $y_i^{t} = y_i^{t-1} - \Delta t \cdot \beta \nabla V(y_i^{t-1}) + \sqrt{2\Delta t} \epsilon_i$,  where $\epsilon_i \sim N(0,1)$}
    \EndFor
    \EndFor
  \end{algorithmic}
\end{algorithm}

\subsection{Exploration Component} 
\label{Sec:exp}

The second stage in each iteration is to collect useful information based on current locations of $\{y_j\}_{j=1}^{\widehat{N}}$. The most useful information related to a new high-probability region is its mode. To this end, we introduce 
the exploration component (EC), a mode finding procedure proposed in \cite{tawn2021annealed} as a component of ALPS algorithm. We adopt the same scheme as ALPS: after $T$ steps of diffusion at tempered level, we apply EC once to locate modes and then decide whether a new mode (or new modes) is found. 

EC consists of two steps: first, it finds the nearest mode $\mu_j$ to $y^t_j$ through local optimization initialized from the current location of $y^t_j$; second, it determines whether $\mu_j$ is a new mode or it has already been contained in $M^t$. Because $N$ could be large and it would be wasteful to run optimization for every particles, the $y^t_j$ will only be a batch of particles uniformly chosen from $Y^t$. The batch size $B$ is a given parameter.  Empirically, $B= 0.01 N$ is used. 

In order to decide whether $\mu_j$ is a new mode, it is necessary to define a distance between modes. One approach is presented in \cite{tawn2021annealed}: 
\begin{equation}
	D(\mu_k,\mu_l):= d^{-1}\max \left\{ (\mu_k-\mu_l)^T \Sigma_k^{-1}(\mu_k-\mu_l), (\mu_k-\mu_l)^T \Sigma_l^{-1}(\mu_k-\mu_l) \right\}
\label{eq:distance}
\end{equation}
where $d$ is the dimension. A threshold $thr$ should be given. If for all $\mu^*$ in $M^t$, $D(\mu_j,\mu^*) > thr$, then $\mu_j$ is considered a new mode. 
Empirically we let $thr = 1 + (2/d)^{1/2}$, as suggested by \cite{tawn2021annealed}.

\begin{algorithm}
  \caption{The exploration component}
    \label{alg:explore}
  \begin{algorithmic}[1]
   	\Require{Particles at hotter temperature $Y^t = \{y^t_j\}_{j=1}^N$, and information of modes $I^{t}:=\{  M^t,S^t,W^t\}$ discovered by iteration $t$; a threshold $thr$, and a batch size $B$.}
    \Procedure{ModeFinder}{$Y^t,I^t$} 
    \State{New = False}
    \State{$\bar{Y} = \{y_{k_1},y_{k_2},...,y_{k_B}\}$, where $k_1,...,k_B$ are uniformly chosen from $\{1,2,...,N\}$}
     \For{$b \gets 1 \textrm{ to } B$}
	\State{\textbf{Run:} Quasi-Newton optimization initialized from $y_b$ to get $\bar{\mu}_b \in \R^d$.}
	\State{Compute: 
	\begin{equation}
		\bar{\Sigma}_b :=- \left(  \nabla^2 \log \pi(\bar{\mu}_b)\right)^{-1} \nonumber
		\end{equation}}
	\For{$k \gets 0 \textrm{ to } m_t$}
		\State{Compute: $d_k:=D(\mu_k,\bar{\mu}_b)$} \Comment{where $D(\cdot,\cdot)$ is defined in \eqref{eq:distance}}
		\EndFor
		\If{$\min \{d_1,\ldots,d_{m_t}\} > thr$ }
		\State{New = True}
		\State{Update $I^t$ by adding the new mode information: 
		\begin{equation}
			(\mu_{m_t+1}, \Sigma_{m_t+1},\widehat{w}_{m_t+1}):=(\bar{\mu}_b, \bar{\Sigma}_b,\bar{w}_b)
		\end{equation}
		}
		\EndIf
		\EndFor
    \EndProcedure
	\State \Return{$I^{t}$, New}
  \end{algorithmic}
\end{algorithm}

\subsection{Within-temperature Updates at Target Level} 
\label{Sec:ALSBD}

\subsubsection{Proposing new modes}
The third stage in each iteration is within-temperature updates of $X$. There are two cases according to the output of EC: \par
1. If a new mode is found by EC, then the information about new mode needs to be passed to $X$. We incorporate the information into Gaussian approximation $\hat{\rho}(\cdot;I^t)$, defined in \eqref{eq:defproposal}. The Gaussian mixture will serve as proposal distribution in Metropolis-Hastings steps. Denote the $N$ proposals at step $t$ as $z^t_1,z^t_2,...,z^t_N \overset{\mathrm{iid}}{\sim} \hat{\rho}(\cdot;I^t)$, the Metropolis-Hastings acceptance probability for each sampler is computed as:
\begin{equation}
A^t_i = \min \biggl\{1, \frac{\hat{\rho}(z^t_i;I^t) \pi(x^t_i)}{\hat{\rho}(x^t_i;I^t) \pi(z^t_i)} \biggr\}, \quad \text{for } i=1,2,...,N.
\end{equation}
Then we assign $x^{t+1}_i = z^t_i$ with probability $A_i^t$, or $x^{t+1}_i = x^t_i$ otherwise. Through the MH step, some of the samplers are transported to the area around new modes, so the new high-probability regions are now covered by $\text{supp}(\rho_{t+1})$. After that, we apply the Langevin algorithm with birth-death process (BDLS) for updates of $X$. From previous literature \cite{lu2019accelerating}, we know that birth-death process can transport particles between separated high-probability regions (see explanation in section~\ref{subsec:BDLS}). Therefore, a bunch of samplers that clustered around old modes will be transported to the new modes with little samplers around. 

2. If no new mode is found, we skip the Metropolis-Hastings steps and only apply BDLS for updates of $X$, as follows. 

\subsubsection{Birth-Death process accelerated Langevin sampling}
\label{subsec:BDLS}
BDLS consists of two parts:  local updates based on the unadjusted Langevin algorithm (see algorithm~\ref{alg:ULA}), and a birth-death process. BDLS was described in detail in \cite{lu2019accelerating} and restated in Algorithm~\ref{alg:BDLS}. 
We choose BDLS because it extends beyond the limitations of local moves. The birth-death process allows the sampler to transition between high probability areas without being hindered by energy barriers.

\paragraph{Pure birth-death process} 
Usually, there are two types of birth-death process, which separately correspond to two types of birth-death dynamics, one is governed by relative entropy \eqref{eq:BD1}, while the other is governed by $\chi^2$-divergence \eqref{eq:BD2}.
\begin{equation}
\label{eq:BD1}
\partial_t \rho_t = - \big(\log \frac{\rho_t}{\pi} 
- \int_{\R^d} \rho_t \log \frac{\rho_t}{\pi} dx \big) \rho_t
\end{equation}
\begin{equation}
\label{eq:BD2}
\partial_t \rho_t = -\big( \frac{\rho_t}{\pi} 
- \int_{\R^d} \frac{\rho_t^2}{\pi} dx \big) \rho_t
\end{equation}
They share the same form of 
\begin{equation}
\label{eq:BD}
\partial_t\rho_t = - \alpha_t \rho_t
\end{equation}
where $\alpha_t$ is called birth-death rate. It is easy to intuitively interpret the birth-death dynamics. $\alpha_t(x) > 0$ implies that $\rho_t$ is relatively large at $x$ compared to $\pi$, which means there are too much particles around $x$, so $\partial_t \rho_t(x) < 0$; vice versa. Theoretically, the birth-death dynamics were also characterized as gradient-flows under the spherical Hellinger distance $d_{\text{SH}}$ : \eqref{eq:BD1} follows a gradient flow of KL-divergence $KL(\rho_t | \pi)$ under $d_{\text{SH}}$, while \eqref{eq:BD2} follows a gradient flow of $\chi^2$-divergence $\chi^2(\rho_t | \pi)$ under $d_{\text{SH}}$. The definition of $d_{\text{SH}}$ and derivation of the gradient flow structure were given in \cite[Section 2,3]{lu2022birth}. \cite{lu2019accelerating} chose the dynamic governed by relative entropy \eqref{eq:BD1}, we use the second birth-death dynamic \eqref{eq:BD2} instead. To implement the birth-death process, we apply the Euler-Maruyama scheme
\begin{equation*}
\rho_{t + \Delta t} = e^{-\alpha_t \Delta t} \rho_t, \text{ where } \alpha_t = \frac{\rho_t}{\pi} - \int_{\R^d} \frac{\rho_t^2}{\pi} dx 
\end{equation*}In order to calculate the birth-death rate $\alpha_t$, we apply kernel density estimation to approximate $\rho_t$, which was already given by \eqref{itemrho}.

\paragraph{Langevin sampling with birth-death process} BDLS is a combination of Langevin sampling with pure birth-death process,
\begin{equation}
\label{eq:dy_BDLS}
    \partial_t \rho_t = \nabla \cdot (\rho_t \nabla V + \nabla \rho_t) - \alpha_t \rho_t
\end{equation}
Its implementation is simple: first apply Langevin diffusion to each particles separately, and then apply birth-death process, see algorithm~\ref{alg:BDLS}.\\
The key observation that has inspired our research is that the BDLS exhibits a notably sluggish exploration rate. This can be understood from its dynamic \eqref{eq:dy_BDLS}. Initially, in pure birth-death dynamic, see \eqref{eq:BD}, we note that $\partial_t \rho_t(x)$ remains zero whenever $\rho_t(x)$ itself is zero, so the support of $\rho_t$ never expand. This implies that the exploration process of BDLS solely relies on the Langevin component, which means its exploration rate cannot surpass that of pure Langevin dynamics. Recall that Langevin sampling has a slow diffusion rate limited by the time step and a slow mixing between separate modes limited by energy barrier, it becomes evident why BDLS as a whole has a slow exploration rate. 

 \begin{algorithm}
  \caption{Birth-death Process Accelerated Langevin Sampling (BDLS)}
    \label{alg:BDLS}
  \begin{algorithmic}[1]
	\Require{$V(x)$, the log-likelihood of target distribution $\pi(x)$; a set of initial particles $\{x^0_i\}_{i=1}^N$; the kernel function $K$; $T$ the number of updates, $\Delta t$ the time step.}
    
	\For{$t \gets 1 \textrm{ to } T$}
	
	\State{\textbf{Langevin:}}
	\For{$i \gets 1 \textrm{ to } N$}
    \State{set $x_i^{t} = x_i^{t-1} - \Delta t \nabla V(x_i^{t-1}) + \sqrt{\Delta t}\epsilon_i$,  where $\epsilon_i \sim N(0,1)$}
    \EndFor
    
    \State{\textbf{Birth-Death process:}}
    \For{$i \gets 1 \textrm{ to } N$}
	\State{calculate $\alpha_i = \big(\frac1N \sum_{l=1}^{N} K(x_i^{t}-x_l^{t})\big) \cdot e^{V(x_i^{t})} $} 
	\EndFor
	
	\State{set $\bar{\alpha}=\frac{1}{N} \sum^N_{l=1}\alpha_l$ }
	
	\For{$i \gets 1 \textrm{ to } N$}
	\If{$\alpha_i > \bar{\alpha}$}
	\State{kill $x_i^t$ with probability $1-e^{(\bar{\alpha}- \alpha_i)\Delta t}$;}
    \State{duplicate one uniformly chosen from other $X^t$;}
	\Else{}
	\State{duplicate $x_i^t$ with probability $1-e^{(\alpha_i-\bar{\alpha})\Delta t}$;}
    \State{kill one uniformly chosen from other $X^t$.}
	\EndIf
	\EndFor
	\EndFor
  \end{algorithmic}
\end{algorithm}

\paragraph{Interacting particle systems}
One interesting aspect of the birth-death process is that it can be interpreted as interaction between $N$ samplers. Therefore, BDLS, as well as pure birth-death process, can be viewed as an interacting particle system, similar to another sampling method called `Ensemble MCMC' recently proposed by Lindsey et al. \cite{lindsey2022ensemble}. Here, we provide a brief comparison between the pure birth-death process \cite{lu2019accelerating} and approach in \cite{lindsey2022ensemble}, which also explains the reason why we take BDLS to adjust weights between separate modes in our method. 

The pure birth-death process is a time discretization of the birth-death dynamic given by \eqref{eq:birthdeath}, similar to the PDE of Ensemble MCMC approach proposed in \cite{lindsey2022ensemble}, as shown in \eqref{ensemble1}, where $\mathcal{Q}$ is a transition operator of its proposal distribution in Metropolis-Hastings. In both cases, the dynamics involve a term that depends on the ratio of the current density $\rho$ to the target density $\pi$, a term that accounts for the interaction between samplers.

\begin{equation}\label{eq:birthdeath}
    \frac{d \rho}{dt} = -(\frac{\rho}{\pi} - z_{\rho}) \rho, z_{\rho} := \int \frac{\rho(x)}{\pi(x)} \rho(x) dx
\end{equation}

\begin{equation}
\label{ensemble1}
\frac{d \rho}{dt} = - \frac{1}{Z_{\rho}} \bigl(\frac{\rho}{\pi} - Z_{\rho}\bigr) \mathcal{Q}\rho, Z_{\rho} := \int \frac{\rho(x)}{\pi(x)} \mathcal{Q}\rho(x) dx
\end{equation}
In practice, the proposal $\mathcal{Q}$ is chosen as a local random walk, for example, Gaussian diffusion. These two dynamics become almost the same under one specific case. When $\mathcal{Q} \to \text{Id}$, \eqref{ensemble1} becomes 
\begin{equation}
    \frac{d \rho}{dt} = -\frac{1}{z_{\rho}} \bigl(\frac{\rho}{\pi} - z_{\rho}\bigr) \rho,
\end{equation}
which differs from \eqref{eq:birthdeath} only in scale. The difference in scale reflects difference in implementation. Although both algorithms need to calculate the rate $r(x) = \frac{\rho(x)}{\pi(x)} - z_{\rho}$ for $N$ samplers in each iteration, the ensemble MCMC method only adjust one particle per iteration, while the birth-death process adjust the whole pack per iteration. Correspondingly, the ensemble MCMC method only applies density estimation for once per iteration. while the birth-death process applies density estimation for $N$ times per iteration. Accordingly, they have the same cost scaling per effective update. 
Both methods facilitate interaction among particles, leveraging global insights to transcend the limitations of local updates. However, while the ensemble MCMC method can attain asymptotic exactness, the BDLS may fall short in this regard.

\section{Theoretical Analysis}

\subsection{Main results}
\label{sec:mainresults}

For the theoretical property of BDEC, we choose to analyze the continuous time sampling dynamic instead of discrete-time Markov Chain, since the analysis of convergence of former is easier and cleaner than that of a discrete process. Here, we only focus on the dynamic of $X^t$ while leaving $Y^t$ in the background, since we are only concerned about how samplers at original temperature converge. Of course, one should keep in mind that the exploration of $Y^t$ plays an important role affecting the convergence of $X^t$.

Let us first present the continuous-time sampling dynamics. Let $\rho_t$ denote the density of samplers $X^t$ at target level by time $t > 0$.  Let $\Omega$ denote the support of $\pi$. We assume $\rho_t \ll \pi$ throughout the whole process, which means $\rho_t$ is also supported inside $\Omega$. 

The distribution evolves according to:
\begin{equation}
\label{dynamicofrho}
\partial_t \rho_t = \nabla \cdot (\rho_t \nabla V + \nabla \rho_t) - \tau_1 \alpha_t \rho_t + \tau_2  \int_{\Omega} R(x,y)(\pi(x)\rho_t(y) - \pi(y) \rho_t(x)) dy.
\end{equation}
The three terms on the right hand side correspond to the following:
\begin{itemize}
    \item The first one is given by the Fokker-Planck equation of Langevin dynamic,
    \begin{equation*}
    \partial_t \rho_t = \nabla \cdot (\rho_t \nabla V + \nabla \rho_t)
    \end{equation*}
    \item The second one $- \tau_1 \alpha_t \rho_t$ stands for the Birth-Death process, with birth/death rate 
    $$\alpha_t = \frac{\rho_t}{\pi} - \int_{\Omega_t} \frac{\rho_t^2}{\pi} dx.$$
    The integral is subtracted in order to keep the normalization of density $\int_{\Omega_t} \rho_t dx = 1$.
    \item The third one stands for the Metropolis-Hastings step after new modes are found (see case 1 in section \ref{Sec:ALSBD}). We use the shorthand notation $\hat{\rho}_t(x)$ for the proposal density $\hat{\rho}(x;I_t)$ which was defined in \eqref{eq:defproposal}. Along pure Metropolis-Hastings, $\rho_t$ evolves according to 
    \begin{align*}
\partial_t \rho_t = & \int_{\Omega} \min \{ \frac{\hat{\rho}_t(x)}{\pi(x)}, \frac{\hat{\rho}_t(y)}{\pi(y)} \} (\pi(x)\rho_t(y) - \pi(y) \rho_t(x)) dy \\
= & \int_{\Omega} \min \{ 1, \frac{\hat{\rho}_t(y)\pi(x)}{\hat{\rho}_t(x)\pi(y)} \} \rho_t(y) \hat{\rho}_t(x) - \min \{ 1, \frac{\hat{\rho}_t(x)\pi(y)}{\hat{\rho}_t(y)\pi(x)} \}\rho_t(x) \hat{\rho}_t(y)dy
\end{align*}
    The term $R(x,y)$ in \eqref{dynamicofrho} is then given by the MH acceptance rate $A(x,y)$ (defined in \eqref{eq:acceptancerate}) multiplying with $\hat{\rho}_t(x)/\pi(x)$:
$$ R(x,y) = \min \biggl\{ \frac{\hat{\rho}_t(x)}{\pi(x)}, \frac{\hat{\rho}_t(y)}{\pi(y)} \biggr\}.$$
One need to remember that, samplers at target level stay still during the mode finding process EC (see \ref{Sec:exp}), but only do Metropolis-Hastings update afterwards. So throughout this section, we call this part of update `MH step' instead of `EC'. 
\end{itemize}
In \eqref{dynamicofrho}, $\tau_1, \tau_2$ are parameters that determine the frequency (relative to the Langevin update) of  Birth-Death step and  Metropolis-Hastings step. 
In our implementation, $\tau_1 = 1$, as a Langevin update is always followed by a birth-death adjustment. In the implementation, the value of $\tau_2$ depends on the frequency a new mode is discovered, and becomes $0$ once all modes are found. However, for simplicity of the analysis, we can also assume $\tau_2 = 1$, which means, one step of MH update is carried out in each iteration. To make clear comparison with Langevin algorithm with birth-death process \cite{lu2019accelerating}, we take $\tau_1 = 1$ and $\tau_2$ to be strictly positive without loss of generality. We will discuss the case $\tau_2 = 0$ in Remark~\ref{rm:tau}. 

Let us define relative density $g_t = \rho_t/\pi$ and $\hat{g}_t = \hat{\rho}_t/\pi$. From \eqref{dynamicofrho} we get:
\begin{equation} \label{dynamicofg}
\partial_t g_t = -\nabla V \cdot \nabla g_t + \Delta g_t - \tau_1 (g_t - \bar{g}_t)g_t + \tau_2 \int_{\Omega} \min(\hat{g}_t(x), \hat{g}_t(y))(g_t(y) - g_t(x))\pi(y)dy,
\end{equation}
where $\bar{g}_t = \int_{\Omega} g_t^2\pi dx$. 
Based on this, we analyze the convergence rate of $\rho_t$ in the sense of $\chi^2-$divergence, 
\begin{equation}
\label{D}
    D_t = \int_{\Omega} \frac{\rho_t^2}{\pi} dx - 1 =  \int_{\Omega} g_t^2 \pi dx - 1,
\end{equation} 
which would converge to $0$ when $\rho_t \to \pi$, i.e., $g_t \to 1$.


To analyze the convergence behavior of our proposed algorithm, it is crucial to evaluate the exploration progress of both the sampling dynamics, $\rho_t$, and the proposal distribution, $\hat{\rho}_t$. Specifically, we introduce the following assumption regarding the proposal distribution:

\begin{assumption}
\label{asp:explored}
    Suppose $\hat{\rho}_t$ is absolutely continuous and there exists a time $t_0 \geq 0$ and a parameter $c > 0$ such that for all $t \geq t_0$,
    \begin{equation}
    \label{eq:assumptionG}
        \inf_{x \in \Omega} \frac{\hat{\rho}_t(x)}{\pi(x)} \geq c.
    \end{equation}
    Under this condition, we say that the exploration component has fully explored the state space at time $t_0$. 
\end{assumption}

This assumption is satisfied when all the modes of the target distribution have been sufficiently explored; in other words, when the ensemble of Markov Chains $Y^t$ have traversed the entire state space $\Omega$. A detailed explanation of this assumption is provided in Section~\ref{sec:asp_c}. It is important to note that $\hat{\rho}_t$ does not represent the law of $Y^t$. Instead, it is a Gaussian mixture constructed using the information gathered by $Y^t$.

Throughout the analysis, we impose an additional condition on $g_t$. \begin{assumption}[Periodic Boundary Condition]
\label{asp:periodic}
Let $g_t$ solves \eqref{dynamicofg} for $t \geq 0$. 
    Suppose $g_t(\cdot): \Omega \to \R^{+}$ satisfies the periodic boundary condition on $\Omega$.  
\end{assumption}

We first establish a result demonstrating that once $Y^t$ has thoroughly explored the entire state space, the process $X^t$ will subsequently achieve full exploration of the state space as well.

\begin{lemma}
\label{lem:G_t_lower_1}
Suppose $\Omega$ a compact set in $\R^d$ and assumption~\ref{asp:periodic} holds. 
    Let $\rho_t$ solve \eqref{dynamicofrho} for $t \geq 0$, and $\tau_1, \tau_2 \geq 0$. 
    If assumption~\ref{asp:explored} holds for $t_0 \geq 0$ and $c > 0$,
    we have 
    \begin{equation}
    \label{eq:inf_g_0}
        \inf_{x \in \Omega} \frac{\rho_t(x)}{\pi(x)} \geq 1 - e^{-c\tau_2 (t - t_0)},
    \end{equation}
    which implies, when $t \geq t_0 - (c\tau_2)^{-1}\log(1 - e^{-M})$, we have 
    \begin{equation}
    \label{eq:inf_g}
        \inf_{x \in \Omega} \frac{\rho_t(x)}{\pi(x)} \geq e^{-M}.
    \end{equation}
\end{lemma}

Note that the lower bound of the relative density in \eqref{eq:inf_g_0} converges to $1$ as $t \to \infty$. When the lower bound reaches $1$, it follows that $\rho_t = \pi$ almost surely. In the following lemma, we demonstrate that the convergence rate can be further improved.

\begin{lemma}
\label{lem:G_t_lower_2}
Suppose $\Omega$ a compact set in $\R^d$ and assumption~\ref{asp:periodic} holds.
Let $\rho_t$ solve \eqref{dynamicofrho} for $t \geq 0$, $\tau_1 = 1$ and $\tau_2 \geq 0$. If there exists a time $t_1 \geq 0$ and $M \geq 0$, such that 
\begin{equation}
\label{eq:inf_g_2}
    \inf_{x \in \Omega} \frac{\rho_{t_1}(x)}{\pi(x)} \geq e^{-M}
\end{equation}
then for any $t \geq t_0$,  
\begin{equation*}
    \inf_{x \in \Omega} \frac{\rho_t(x)}{\pi(x)} \geq \frac{1}{e^{-(t - t_1)} (e^M - 1) + 1}. 
\end{equation*}
\end{lemma}

We now present our main theorem, which establishes the convergence of $\rho_t$ in terms of the $\chi^2$-divergence. 

\begin{theorem}
\label{Theorem_main}
Suppose $\Omega$ a compact set in $\R^d$ and assumption~\ref{asp:periodic} holds.
Assume assumption~\ref{asp:explored} holds for $t_0 \geq 0$ and $c > 0$. 
    Let $\rho_t$ solve (\ref{dynamicofrho}) for $t \geq t_0, \tau_1 = 1, \tau_2 > 0$, then $D_t$ is decreasing with exponential rate, 
    \begin{equation}
        D_t \leq D_{t_0} \exp\{ - 2 (\tau_2 c + 1) (t - t_0) - \frac{2}{\tau_2 c} e^{- \tau_2 c (t - t_0)}\} 
    \end{equation}
    Furthermore, for any $ \delta \in (0, 1 - \tau_2 c)$, 
    \begin{equation}
    \label{eq:The4.1}
        D_t \leq D_{t_{*}} \exp \{ - 2 (\tau_2 c + 1 - \delta) (t - t_{*})\}
    \end{equation}
    holds for any $t \geq t_{*} = t_0 - \log \delta + \frac{1}{\tau_2 c}\log\frac{1}{\tau_2 c} - (\frac{1}{\tau_2 c} - 1) \log(\frac{1}{\tau_2 c} - 1)$. 
\end{theorem}

Theorem~\ref{Theorem_main} asserts that, provided Assumption~\ref{asp:explored} holds, $\rho_t$ converges to $\pi$ at a rate independent of the potential function $V$ after a short waiting time. 

The significance of incorporating the exploration component is evident when comparing Theorem~\ref{Theorem_main} to \cite[Theorem 3.3]{lu2019accelerating}, which analyzes the convergence of BDLS. Both methods—BDEC and BDLS—achieve an asymptotic exponential convergence rate that is independent of $V$. However, while the limit convergence rate for BDLS is $2$, our method, BDEC, attains a superior rate of $2 + 2 \tau_2 c$. Moreover, the asymptotic convergence of BDEC relies on weaker assumptions.

In \cite[Theorem 3.3]{lu2019accelerating}, asymptotic convergence is achieved only after condition \eqref{eq:inf_g} (or,  \eqref{eq:inf_g_2}) is satisfied, implying that the current distribution $\rho_t$ must have fully explored the entire state space. The time required for this is unclear and could be significant, especially when $V$ exhibits high energy barrier. In contrast, Lemma~\ref{lem:G_t_lower_1} shows that under Assumption~\ref{asp:explored}, condition \eqref{eq:inf_g} can be satisfied much more quickly.

Notably, assuming a positive lower bound for the relative density $\hat{\rho}_t / \pi$ is a weaker requirement than assuming a positive lower bound for $\rho_t / \pi$. This is because exploration at a higher temperature level typically occurs faster than at the original temperature level. Therefore, the comparison between Theorem~\ref{Theorem_main} and \cite[Theorem 3.3]{lu2019accelerating} highlights the role of the exploration component in accelerating the convergence of Langevin sampling with the birth-death process.

The significance of birth-death process is reflected through the fact that the asymptotic convergence rate is independent of $V$, as explained in \cite{lu2019accelerating}. Our theoretical analysis aligns with the intuitive understanding of the interplay between the exploration component and the birth-death process. The exploration component accelerates the process of $\rho_t$ exploring the state space, while the birth-death process promptly adjusts the weights among all modes once the exploration is complete.

\subsection{Proof of Main Theorem} 
The proofs of Lemma~\ref{lem:G_t_lower_1} and Lemma~\ref{lem:G_t_lower_2} are essentially adaptations of the maximum principle. For the sake of completeness, we include them in Appendix~\ref{ap_sec:proof_lem}. Below, we provide the proof of our main result, Theorem~\ref{Theorem_main}.
\begin{proof} 
\label{proof_main}
We restate some notations so that it can be easier for readers to keep up. $\rho_t$ refers to the law of $X^t$, and $\hat{\rho}_t$ refers to the proposal distribution in its Metropolis-Hastings step. $g_t(x) = \rho_t(x)/\pi(x), \hat{g}_t(x) = \hat{\rho}_t(x)/\pi(x)$. Further, for sake of convenience, we denote the lower bound for $g_t$ as $G_t$, 
\begin{equation}
\label{eq:def_G_t}
    G_t := \inf_{x \in \Omega} g_t(x) = \inf_{x \in \Omega} \frac{\rho_t(x)}{\pi(x)}
\end{equation}
Take the derivative of the $\chi^2$ divergence $D_t = \int_{\Omega} g_t^2 \pi dx - 1$ under $\tau_1 = 1$, we have
\begin{equation}
\label{eq:dD}
\begin{aligned}
\frac{d D_t}{dt} &=  2\int_{\Omega} g_t(x) \partial_t g_t(x) \pi(x) dx \\
& =  2 \int_{\Omega} g_t(x) \nabla\cdot(\pi(x)\nabla g_t(x))dx - 2 \int_{\Omega} (g_t - \bar{g}_t){g_t}^2\pi dx \\
&\quad  + 2\tau_2 \int_{\Omega^2} \min(\hat{g}_t(x), \hat{g}_t(y))(g_t(y) - g_t(x))g_t(x)\pi(x)\pi(y)dydx \\
& =  - 2 \int_{\Omega} \pi(x) (\nabla g(x))^2 dx - 2 \int_{\Omega} (g_t - \bar{g}_t){g_t}^2\pi dx \\
&\quad  + 2\tau_2 \int_{\Omega^2} \min(\hat{g}_t(x), \hat{g}_t(y))(g_t(y) - g_t(x))g_t(x)\pi(x)\pi(y)dydx 
\end{aligned}
\end{equation}
The last equality is generated by using the Gauss-Green formula $\int_{\Omega} \nabla\cdot(\pi g_t \nabla g_t)dx = \int_{\partial \Omega} \pi g_t \nabla g_t \cdot d \vec{S}$. Notice that $\rho_t(x) \vert_{\partial \Omega} = 0$, so $$\int_{\Omega} g_t(x) \nabla\cdot(\pi(x)\nabla g_t(x))dx + \int_{\Omega} \pi (\nabla g_t)^2 dx = 0.$$ 
 
Then we upper bound the right hand side of \eqref{eq:dD}:

1. The first term $- 2 \int_{\Omega_t} \pi (\nabla g_t)^2 dx \leq 0$. 

2. In the second term, $\bar{g}_t$ is defined to be the weighted average of $g_t$ on $\Omega$, 
\begin{equation*}
    \bar{g}_t(x) = \int_{\Omega} g_t^2 \pi dx = D_t + 1
\end{equation*} 
It is easy to verify that 
\begin{equation*}
\label{eq:sec_term_b0}
    - 2 \int_{\Omega}\left(g_{t}-\bar{g}_{t}\right) g_{t}^{2} \pi d x = - 2 \int_{\Omega}\left(g_{t}-\bar{g}_{t}\right)^2 g_{t} \pi d x \leq 0
\end{equation*}
Apply Cauchy Schwartz, 
\begin{equation}
\label{eq:cauchy_ineq}
 \int_{\Omega}\left(g_{t}-\bar{g}_{t}\right)^2 \rho_t d x \cdot
    \int_{\Omega} \frac{(\rho_t - \pi)^2}{\rho_t} dx \geq
    \left( \int_{\Omega} (g_t - \bar{g}_t) (\rho_t - \pi) dx\right)^2   
\end{equation}
The right hand side 
\begin{equation*}
\label{eq:cauchy_ineq_right}
\int_{\Omega} (g_t - \bar{g}_t) (\rho_t - \pi) dx  
 = \bar{g}_t - 1, 
\end{equation*} 
while the term on the left hand side
\begin{equation*}
\label{eq:cauchy_ineq_left}
\int_{\Omega} \frac{(\rho_t - \pi)^2}{\rho_t} dx 
\leq \frac{1}{G_t}\int_{\Omega} \frac{(\rho_t - \pi)^2}{\pi} dx
= \frac{1}{G_t} (\bar{g}_t - 1), 
\end{equation*}
where the first inequality applies the definition of $G_t$ \eqref{eq:def_G_t}. 
Take back to \eqref{eq:cauchy_ineq}, we have 
\begin{equation*}
    \int_{\Omega}\left(g_{t}-\bar{g}_{t}\right)^2 g_t \pi d x \geq G_t (\bar{g}_t - 1) = G_t D_t
\end{equation*}

3. The third term 
\begin{equation} \label{eq:thirdterm}
\begin{aligned}
& 2 \tau_2\int_{\Omega^2} \min(\hat{g}_t(x), \hat{g}_t(y))(g_t(y) - g_t(x))g_t(x)\pi(x)\pi(y)dydx \\
 = & - 2 \tau_2\int_{\Omega^2} \min(\hat{g}_t(x), \hat{g}_t(y))(g_t(y) - g_t(x))g_t(y)\pi(y)\pi(x)dydx \\
 = & - \tau_2 \int_{\Omega^2}\min(\hat{g}_t(x), \hat{g}_t(y)) (g_t(y) - g_t(x))^2\pi(x) \pi(y) dxdy \\
 \leq & - \tau_2 c \int_{\Omega^2} (g_t(y) - g_t(x))^2\pi(x) \pi(y) dxdy \\
 = & - 2 \tau_2 c D_t, 
\end{aligned}
\end{equation}
where the inequality applies again \eqref{eq:assumptionG}. 
To summarize, we have 
\begin{equation}
    \frac{dD_t}{dt} \leq - 2 ( \tau_2 c + G_t)D_t, 
\end{equation} 
If we apply Lemma~\ref{lem:G_t_lower_1} that $G_t \geq 1 - e^{-c\tau_2 (t - t_0)}$, we have
\begin{equation*}
    D_t \leq D_{t_0} \exp\{ - 2 (\tau_2 c + 1) (t - t_0) - \frac{2}{\tau_2 c} e^{- \tau_2 c (t - t_0)}\}.
\end{equation*}
Furthermore, Lemma~\ref{lem:G_t_lower_1} implies for any $M > 0$, there exists a time $t_1 = t_0 - (c\tau_2)^{-1}\log(1 - e^{-M})$ that 
\begin{equation*}
    G_{t_1} = \inf_{x \in \Omega} \frac{\rho_{t_1}(x)}{ \pi(x)} \geq e^{-M}
\end{equation*}
Based on this, we apply Lemma~\ref{lem:G_t_lower_2}: for any $t \geq t_1$, 
\begin{equation*}
   G_t \geq \frac{1}{e^{-(t - t_1)} (e^M - 1) + 1 }.
\end{equation*}
Therefore, if we take $t_{*} = t_1 + \log(e^M - 1) - \log \delta$, we have for any $t \geq t_{*}$, $G_t \geq 1 - \delta$. Accordingly, 
for $t \geq t_{*}$, the equation \eqref{eq:The4.1} holds. Notice $t_{*}$ here is 
\begin{equation*}
    t_{*} = t_0 - \frac{1}{\tau_2 c}\log(1 - e^{-M}) + \log(e^M - 1) - \log \delta 
\end{equation*}
for an arbitrary $M, \delta$ satisfying $1 - \delta \geq e^{-M}$. It reaches minimum at $M = - \log(c\tau_2)$. Taking back to the expression of $t_{*}$, we have 
the result in Theorem~\ref{Theorem_main}.
\end{proof}

\begin{remark}
\label{rm:tau}
    In our result, we let $\tau_1 = 1, \tau_2 > 0$. It is slightly different from our implementation that we set $\tau_2 = 0$ if we know $\rho_t$ has covered the entire state space. However, this is not a big deal. When $\tau_2 = 0$, our scheme recovers exactly BDLS. Our theoretical result also recovers that of BDLS in \cite{lu2019accelerating}. $\rho_t$ has covered the entire state space means \eqref{eq:inf_g_2} is satisfied for a time $t$ and an positive parameter $M$. We can still apply Lemma~\ref{lem:G_t_lower_2}. Then we have for any $ \delta \in (0, 1 - e^{-M})$, take $t_{*} = t_1 + \log(e^{M} - 1) - \log \delta$, 
    \begin{equation*}
        D_t \leq D_{t_{*}} \exp \{ - 2 (1 - \delta) (t - t_{*}) \}.
    \end{equation*}
    Which is the same result as in the statement of \cite[Theorem 3.3]{lu2019accelerating}.
\end{remark}

\subsection{Further discussion}
\label{sec:asp_c}

To make our theoretical analysis complete, we need to show that the assumption~\ref{asp:explored} is reasonable. Specifically, this means showing that $\hat{\rho}/\pi$ can be lower bounded by a positive value. Intuitively, we anticipate that the more accurate the Gaussian approximation is, the higher the positive lower bound we can achieve.

Let us start from the uni-modal case: assume $\pi(x) \sim \exp(-V(x))$ has bounded support inside $\B_2(\mu, R)$, a ball centered at $\mu$ with radius $R$ under $\mathcal{L}^2$ norm, and $\mu$ denotes the unique mode of $\pi$. Let $\Sigma = - [\nabla^2 \log(\pi(\mu))]^{-1}$,  we use $\hat{\rho}(x) := \mathcal{N}(\mu, \Sigma)$ to approximate $\pi$. We intend to give a lower bound for 
$$
\frac{\hat{\rho}(x)}{\pi(x)} = \frac{Z_{\pi} }{\vert\Sigma\vert^{1/2} (2\pi)^{d/2} } \exp \left( V(x) - \frac{(x-\mu)^T \Sigma^{-1} (x-\mu)}{2} \right)
$$
Let us consider a special case where $\pi$ is log-concave with parameter $M$ and log-smooth with parameter $L$, i.e. 
$$
    \frac{M}{2} \|x-y\|_2^2 \leq V(y) - V(x) - \nabla V(x)^{T}(y-x) \leq \frac{L}{2} \|x-y\|_2^2
$$
Without loss of generality we can assume $\mu = 0$ and $V(\mu) = 0$. Then we have 
\[ 
\frac{M}{2} \|x-\mu\|_2^2 \leq  V(x) \leq \frac{L}{2} \|x-\mu\|_2^2 \\
\]
First, from $\Sigma^{-1} = \nabla^2 V(\mu)$, we have 
\[
M \mathbf{I}_d \preceq \Sigma^{-1} \preceq L \mathbf{I}_d \Longrightarrow 
\vert \Sigma \vert^{\frac{1}{2}} \leq M^{\frac{d}{2}}
\]
Then, we lower bound the normalizing constant of $\pi$:
\[
Z_{\pi}= \int_{\B_2(\mu, R)} \exp(-V(x))dx \geq \int_{\B_2(R)} \exp(-\frac{L}{2} \|x\|_2^2)dx
\]
For normal vector $X \in \R^d$ with zero mean and covariance $\Sigma = L^{-1} \mathbf{I}_d$, we have $\mathbb{E}\| X \|_2 \leq \sqrt{d/L}$. We can use Gaussian concentration inequality to obtain a tail bound:
\[
\begin{aligned}
& \Pr[\| X \|_2 \geq \mathbb{E}\| X \|_2 + t] \leq \exp \left( -\frac{L t^2}{2} \right), \forall t \geq 0\\
\Longrightarrow & \Pr[\| X \|_2 \geq \sqrt{d/L}+ t] \leq \exp \left(-\frac{L t^2}{2} \right), \forall t \geq 0 
\end{aligned} \]
Let $t = R -\sqrt{d/L}$, then
\begin{equation}
\label{eq:boundnc}
\begin{aligned}
    Z_{\pi} & = \frac{(2\pi)^{d/2}}{L^{d/2}} \left( 1 - \Pr[\| X \|_2 \geq R] \right) \\
& \geq \frac{(2\pi)^{d/2}}{L^{d/2}} \left[1 - \exp \left(-\frac{L}{2}(R - \sqrt{d/L})^2\right) \right], \forall R \geq \sqrt{d/L}
\end{aligned}
\end{equation}
Finally, in the uni-modal case, we can lower bound the fraction by
\begin{equation}\label{eq:unimodal}
\begin{aligned}
    \frac{\hat{\rho}(x)}{\pi(x)} & = \frac{Z_{\pi} }{\vert\Sigma\vert^{1/2} (2\pi)^{d/2} } \exp \left( V(x) - \frac{(x-\mu)^T \Sigma^{-1} (x-\mu)}{2} \right) \\
& \geq \left( \frac{M}{L} \right)^{d/2} \left[1 - \exp \left(-\frac{L}{2}(R - \sqrt{d/L})^2\right) \right] \exp \left( -\frac{(L-M)R^2}{2}  \right) 
\end{aligned} 
\end{equation}
We can easily extend this result to multimodal case with some mild assumptions. Suppose $\pi$ has $M$ modes and can be expressed by 
\begin{equation} \label{eq:multimodalpi}
    \pi(x) = \sum_{i=1}^M w_i \pi_i(x), \text{with} \sum_{i=1}^M w_i = 1, \text{supp}(\pi) \subseteq \B_2(0, \frac{R}{2})
\end{equation}
where $\pi_i = \exp(-V_i(x))/Z_i$ is probability defined on $\B_2(\mu_i, R)$ with unique mode $\mu_i$. 
We use the Gaussian mixture $\hat{\rho}(x)$ defined in \eqref{eq:defproposal} to approximate $\pi$. Notice that the support of $\pi$ is contained in union of support of each $\pi_i$, 
\begin{equation} \label{eq:multimodal}
   \frac{\hat{\rho}(x)}{\pi(x)} = \frac{\sum_{i=1}^M \widehat{w}_i \mathcal{N}(x; \mu_i, \Sigma_i)}{ \sum_{i=1}^M w_i \pi_i(x)} \geq \min_{i \in [M]} \Big\{  \frac{\widehat{w}_i \mathcal{N}(x; \mu_i, \Sigma_i)}{w_i \pi_i(x)} \Big\} 
\end{equation}
It is sufficient to give a lower bound for each component on the right hand side. Firstly, assume that each $\pi_i$ is $M_i-$convex and $L_i-$smooth, then we can give a bound for $Z_i$ as in \eqref{eq:boundnc}. Secondly, assume that the modes are well-separated, that is, $\pi(\mu_i) \approx w_i \pi_i(\mu_i)$. Then, recall the definition of estimated weights \eqref{eq:defweight}, 
\[
\widehat{w}_i = \frac{w_i \vert \Sigma_i \vert^{1/2} Z_i^{-1}}{\sum_{j=1}^M w_j \vert \Sigma_j \vert^{1/2} Z_j^{-1}}
\]
The denominator 
\begin{equation*}
\begin{aligned}
    & \sum_{j=1}^M w_j \frac{\vert \Sigma_j \vert^{1/2}}{Z_j} \leq \sum_{j=1}^M w_j \cdot \max_{j \in [M]} \frac{\vert \Sigma_j \vert^{1/2}}{Z_j} 
\leq \max_{j \in [M]} \left( \frac{\kappa_j}{2\pi} \right)^{d/2} \left[1 - \Phi(R; L_j) \right]^{-1} \\
& \text{where } \Phi(R; L_j) = \Pr[\|X\|_2 \geq R - \sqrt{d / L_j}], X \sim \mathcal{N}(0, L_j \mathbf{I}_d)
\end{aligned}
\end{equation*}
where $\kappa_j = L_j/M_j$ is condition number for the $j^{th}$ mode, and $\Phi(\cdot)$ is a cumulative distribution function.
Therefore 
\[
\begin{aligned}
\frac{\widehat{w}_i \mathcal{N}(x; \mu_i, \Sigma_i)}{w_i \pi_i(x)}
& = \frac{1}{(2\pi)^{d/2}\sum_{j=1}^M w_j \vert \Sigma_j \vert^{1/2} Z_j^{-1}}
\exp \left( V_i(x) - \frac{(x-\mu)^T \Sigma_i^{-1} (x-\mu)}{2} \right) \\
& \geq \min_{j \in [M]} \Big\{ \kappa_j^{-d/2}\left[1 - \Phi(R; L_j) \right] \Big\} \cdot \exp(- (L_i - M_i)R^2/2)
\end{aligned}
\]
By applying \eqref{eq:defweight}, we obtain a theoretical result concerning Gaussian approximation to multimodal distribution:
\begin{theorem}  
\label{Theorem2}
The target distribution $\pi$ is defined in \eqref{eq:multimodalpi}, and the approximation distribution $\hat{\rho}$ is defined in \eqref{eq:defproposal}. Suppose $V_i$ is $M_i$-strongly convex and $L_i$-smooth, then 
\begin{equation}
\label{eq:Theorem5}
    \inf_{\Omega} \frac{\hat{\rho}(x)}{\pi(x)} \geq \min_{j \in [M]} \biggl\{ \left( \frac{M_j}{L_j} \right)^{d/2}\left[1 - \Phi(R; L_j) \right] \biggr\} \cdot \min_{i \in [M]} \exp(- (L_i - M_i)R^2/2)
\end{equation}
holds, where $\Phi(R; L_j)$ can be upper bounded by $\exp \left(-L_j(R - \sqrt{d/L_j})^2/2\right)$.
\end{theorem}

Theorem \ref{Theorem2} draws a picture about how good or bad the Gaussian approximation can be, under some mild assumptions that takes multimodality into account. It gives 
a positive lower bound for $\widehat{G}_t$, under the condition that all modes have been discovered in the tempered level by time $t$.
If we let $M = 1$, we directly get: 

\begin{corollary}
    Target distribution $\pi \sim \exp(-V)$, with unique mode $\mu$, is defined on $\Omega \subseteq \mathbb{B}_2(\mu,R)$. If $V$ is $M-$strongly convex and $L-$smooth, then \eqref{eq:unimodal} holds for any $x \in \Omega$.
\end{corollary}

\begin{remark}
    Denote the right hand side of \eqref{eq:unimodal} as a function $C(d, M, L, R)$. This lower bound decreases when the condition number $\kappa = L/M$ increases or the dimension $d$ increases. It is under the curse of dimension. Then one may wonder if this lower bound can be improved. However, this lower bound is sharp, because we can construct a function $V(x)$ that has Hessian $\nabla^2 V(\mu) = M \mathbf{I}_d$ locally, while $V(x) = L \| x- \mu \|_2 /2$ for all $x \not\in \B(\mu, \epsilon)$ for a $\epsilon$ arbitrarily small. Under this case, all inequalities we used becomes equation within $\epsilon$ error.
\end{remark}

\begin{remark}
    It is worth noticing that $\exp(-(L-M)R^2/2)$ goes to $0$ as $R \to \infty$. This observation suggests that if $\pi$ is unbounded, the lower bound $c$ of $\widehat{G}_t$ would be arbitrarily small, unless our approximation match the tail behavior of the target distribution. Initially, this may seem concerning, given the near-impossibility of accurately matching the tail behavior of a non-trivial distribution with a Gaussian approximation. However, upon closer examination from a broader perspective, this issue is not as problematic as it appears. Remember that the bound $c$ is employed only in integrals with the form $\int_{\Omega} f(x) \pi(x) dx$, see \eqref{eq:thirdterm}, where low-probability regions make negligible contributions. Accordingly, we can do approximation $\int_{\B(0, R)} f(x)\pi(x) dx \approx \int_{\Omega} f(x)\pi(x) dx$, which means it is reasonable to restrict ourselves to bounded case. the case when $\pi$ is a combination of bounded $\pi_i$ \eqref{eq:multimodalpi}, and one should not be worried about matching the tail behavior. 
\end{remark}

\section{Numerical experiments}

In numerical experiments, we compare the efficiency of our algorithm with BDLS and ALPS by using the same settings as in their original experiments. In order to highlight the importance of birth-death process, we also construct a degraded sampling scheme LEC, which stands for Langevin sampling with exploration component. LEC differs from BDEC only in a way that LEC does not perform the birth-death process. \par
In BDEC algorithm, we keep $N$ Markov chains instead of recording the trajectory of a single Markov chain as \cite{tawn2021annealed} did. These chains are independent while doing within-temperature updates, and they interact during birth-death process. So in order to compare fairly, the ALPS algorithm we use is slightly different from its original version. We construct $N$ independent Markov chains at each temperature level, update them according to the ALPS scheme and we do not record their trajectory. One iteration in ALPS contains $T-1$ within-temperature updates and 1 swap move, which corresponds to $T$ within-temperature updates in BDEC. 
In our kernel density estimation \eqref{itemrho}, 
we choose kernel to be Gaussian kernel $K(x,y) = \frac{1}{(2\pi h^2)^{d/2}} \exp(-|x-y|^2/2h^2)$. In our examples, we always choose $h= 0.05$, but it needs some tuning in other settings.\par
All experiments are conducted using Python
3.8.5 on a personal laptop with a 2 GHz Quad-Core Intel Core i5 processor.
In each example, we run the algorithms for 10 times and average the results.

\subsection{Example 1: 2D Gaussian mixture}
\label{setting1}

\begin{figure}[tbhp]
\centering
\subfloat[Estimating $\mathbb{E}(|x|)$]{\label{fig:201}\includegraphics[width=0.333\textwidth]{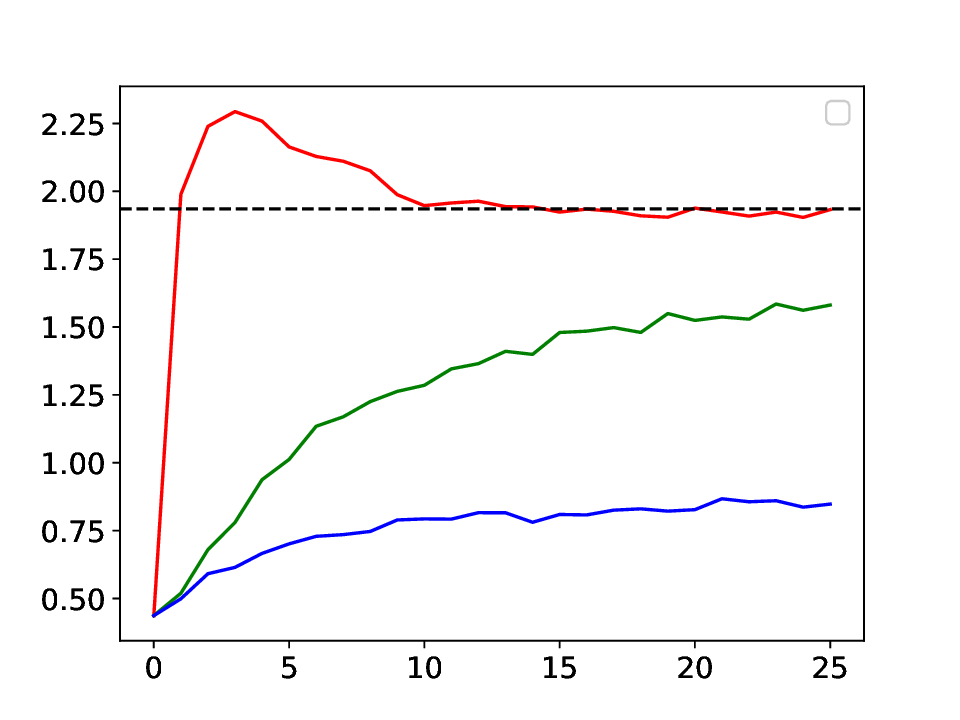}}
\subfloat[Estimating $\mathbb{E}(y)$]{\label{fig:202}\includegraphics[width=0.333\textwidth]{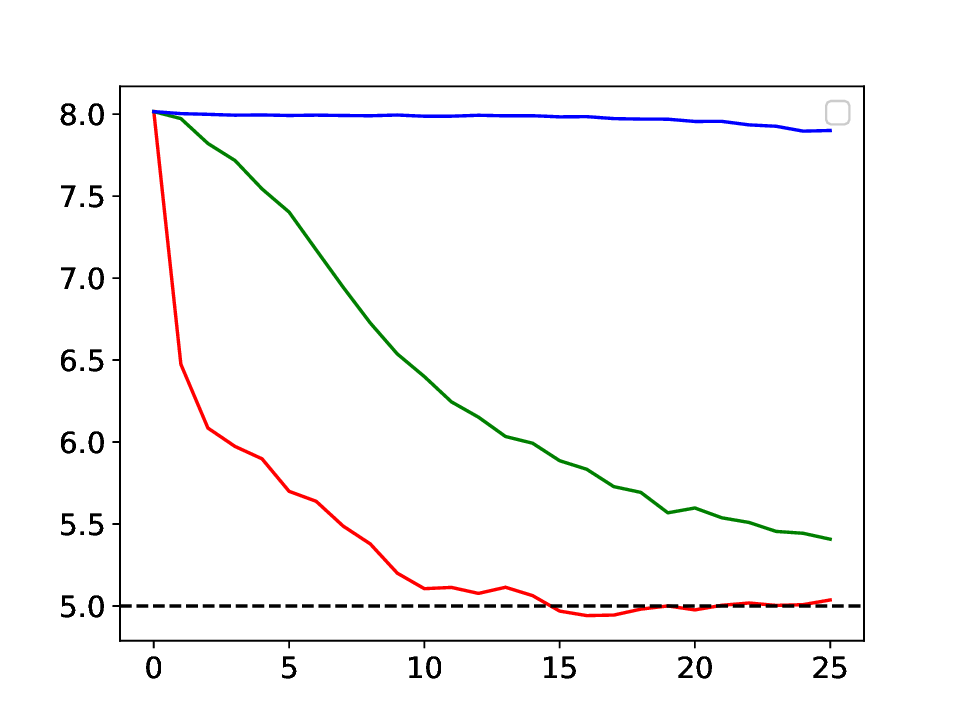}}
\subfloat[Estimating $\mathbb{E}(x^2/3+y^2/5)$]{\label{fig:203}\includegraphics[width=0.333\textwidth]{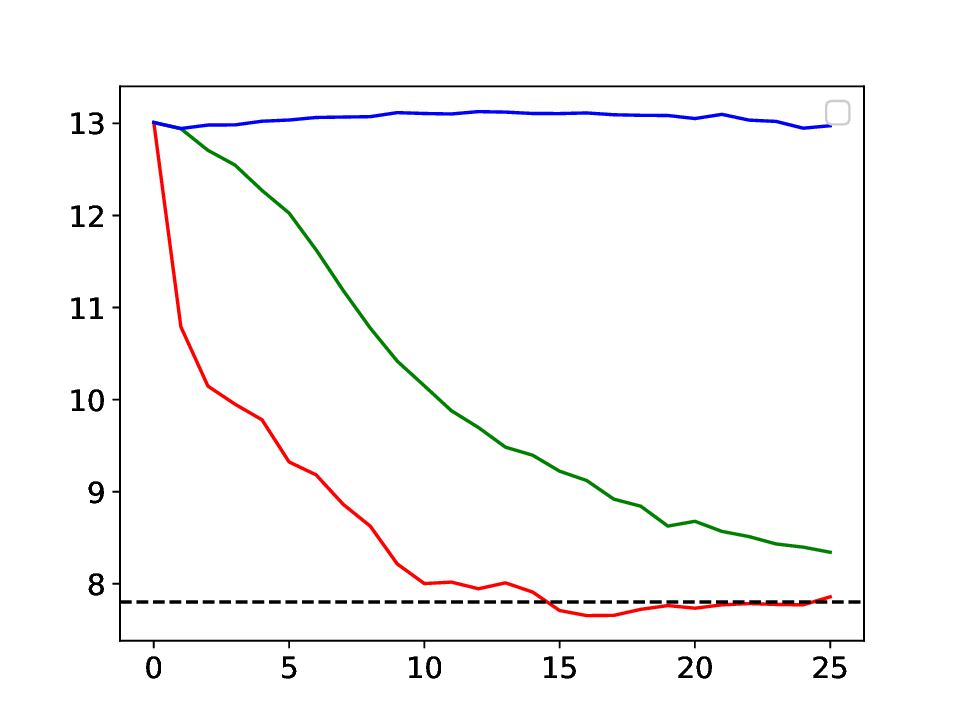}}
\caption{Results of estimating $\mathbb{E}(f)$ for three different $f$ in Example 1. Red, estimation of BDEC; Green, ALPS; Blue, BDLS; Black, the true expectation value. We plot the estimations for 25 total iterations with each iteration including 4 updates.}
\label{fig:example2}
\end{figure}

In order to compare BDEC with the BDLS algorithm, we experiment on a 2-dimensional model from \cite{lu2019accelerating}. The target distribution is a mixture of four well-separated and heterogeneous 2-dimensional Gaussian distributions:
$$
 \pi(\bm{x}) = \sum_{i=1}^4 w_i \mathcal{N}(\bm{x};\mu_i,\Sigma_i)
$$
We keep $1000$ particles and initialize them from a Gaussian distribution $\mathcal{N}(\bm{x};\mu_0,\Sigma_0)$. The parameters are given by 
$$
\begin{aligned}
& w_i = \frac{1}{4},i=1,...,4, 
\mu_0 = \mu_1 = (0,8)^T,\mu_2 = (0,2)^T,\mu_3 = (-3,5)^T, \mu_4 = (3,5)^T \\
& \Sigma_0 = \left(\begin{matrix}0.3&0\\0&0.01  \end{matrix}\right), \Sigma_1 = \Sigma_2 = \left(\begin{matrix} 1.2&0\\0&0.01 \end{matrix}\right),\Sigma_3 = \Sigma_4 = \left(\begin{matrix} 
 0.01&0\\0&2 \end{matrix}\right)
\end{aligned}
$$
We set time step $\Delta t = 0.005$ and kernel width $h = 0.05$ for both sampling schemes. We choose number of total iterations $J = 25$ and number of within-temperature moves $T = 4$ for BDEC, which equals $100$ updates in BDLS. We choose hotter temperature $\beta_{hot} = 0.05$ for Markov chain $Y$ and a batch size $B = 12$ for the Exploration Component.
In ALPS, we choose the annealing scheme to be $(1,3,9)$.
We compare the numerical results of estimating $\mathbb{E}(f)$ for different $f$ in figure~\ref{fig:example2}, which shows that EC does accelerate the convergence of BDLS. 

\begin{figure}[h!] 
\centering

\includegraphics[width=0.19\textwidth]{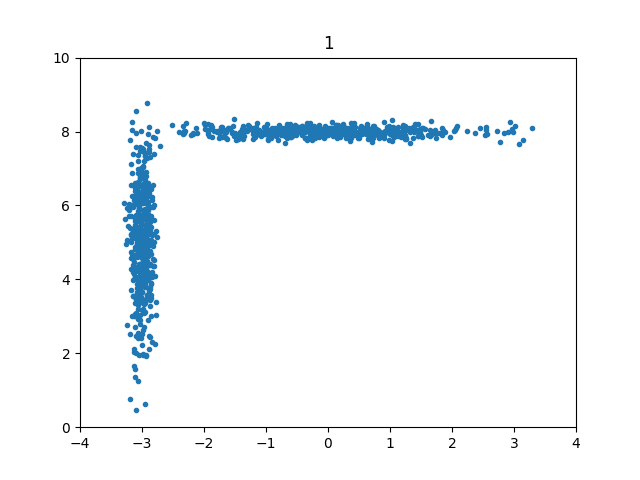}
\includegraphics[width=0.19\textwidth]{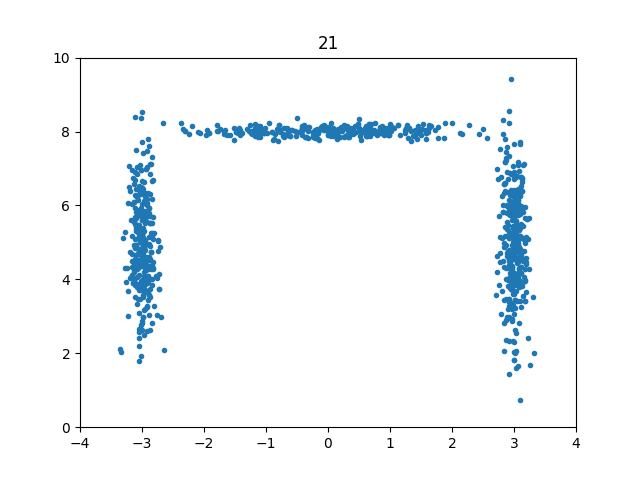}
\includegraphics[width=0.19\textwidth]{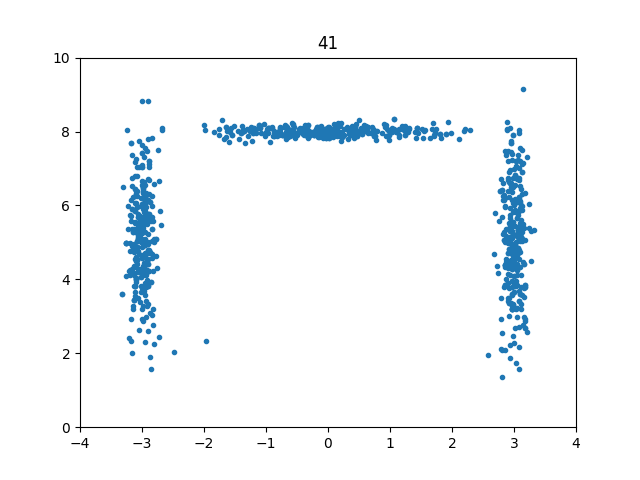}
\includegraphics[width=0.19\textwidth]{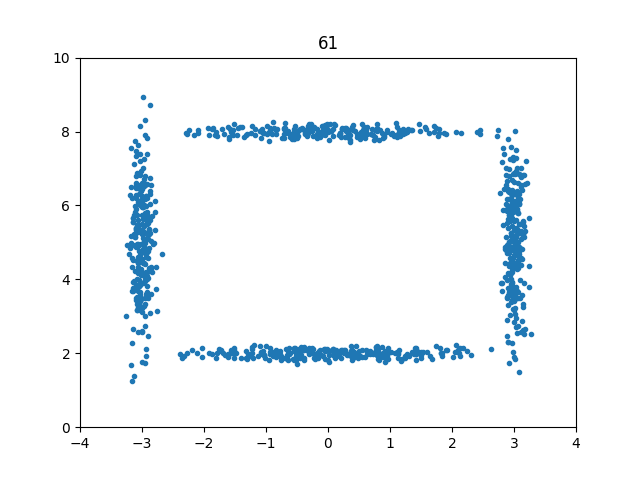}
\includegraphics[width=0.19\textwidth]{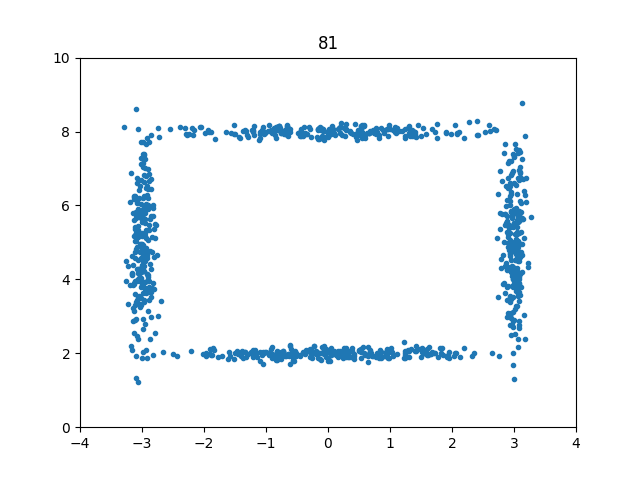}

\includegraphics[width=0.19\textwidth]{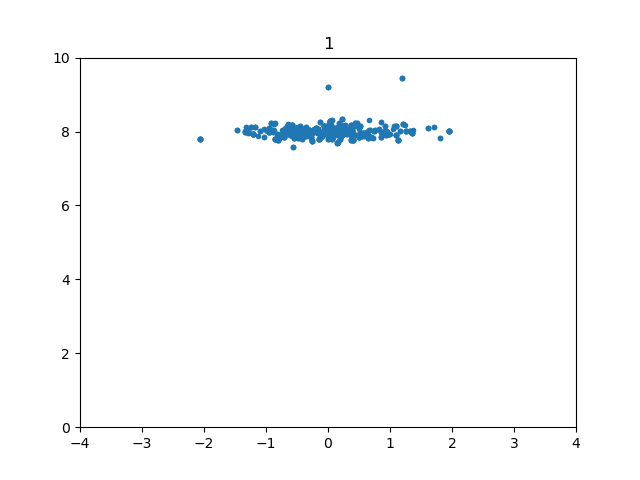}
\includegraphics[width=0.19\textwidth]{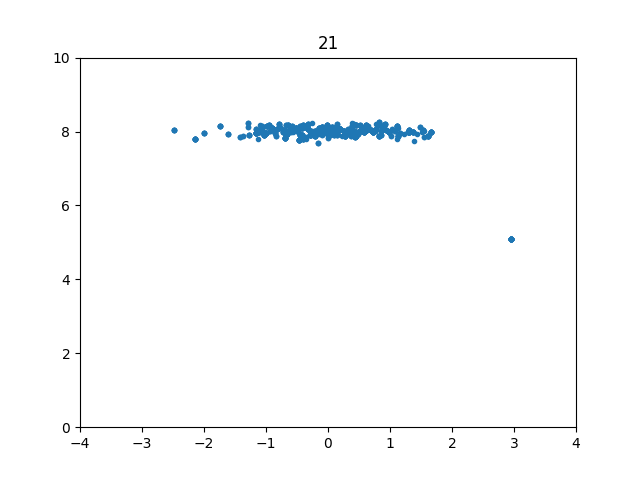}
\includegraphics[width=0.19\textwidth]{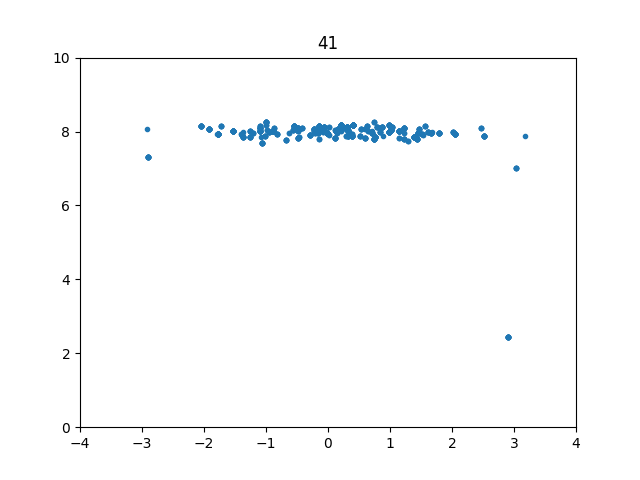}
\includegraphics[width=0.19\textwidth]{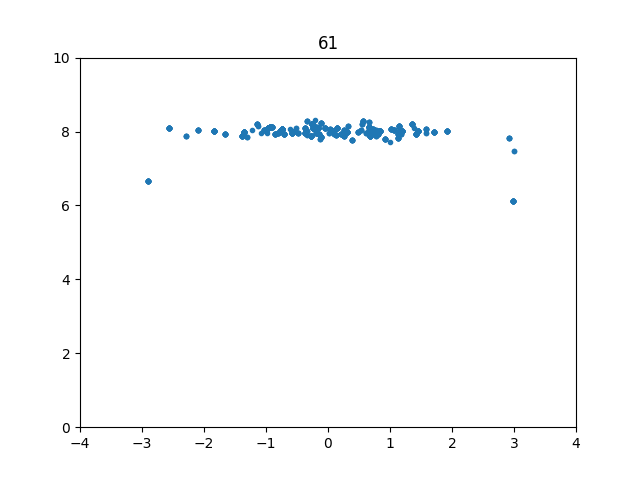}
\includegraphics[width=0.19\textwidth]{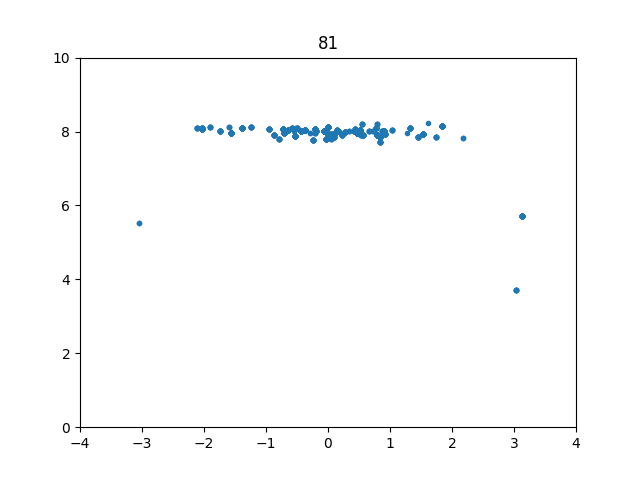}

\includegraphics[width=0.19\textwidth]{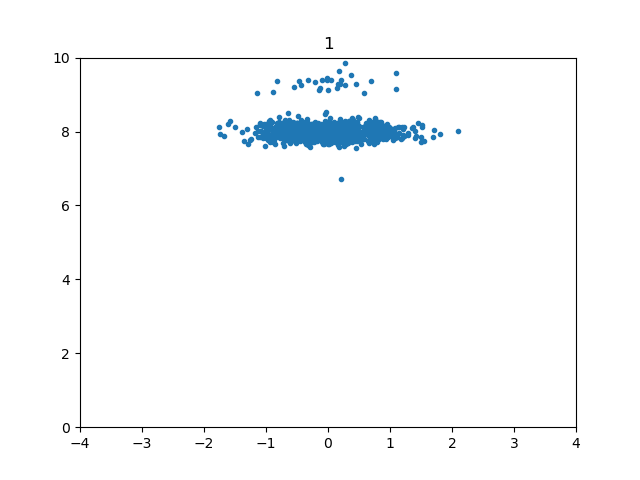}
\includegraphics[width=0.19\textwidth]{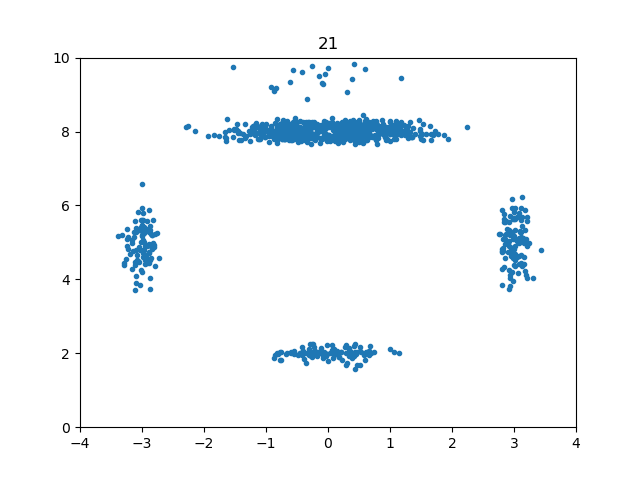}
\includegraphics[width=0.19\textwidth]{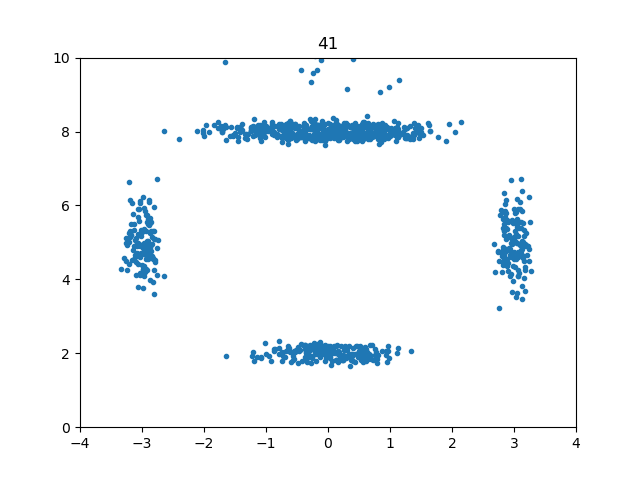}
\includegraphics[width=0.19\textwidth]{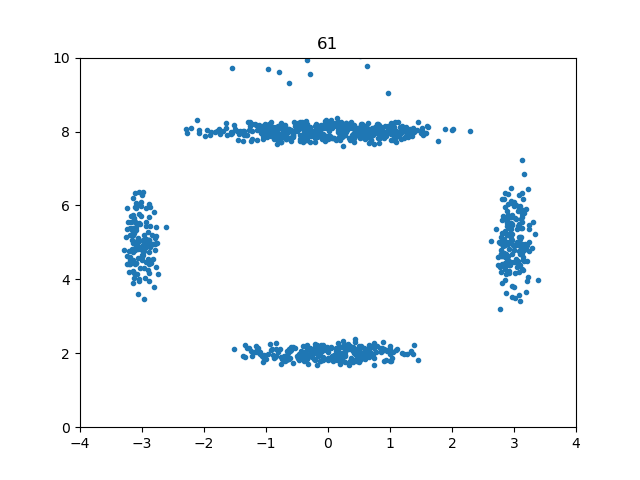}
\includegraphics[width=0.19\textwidth]{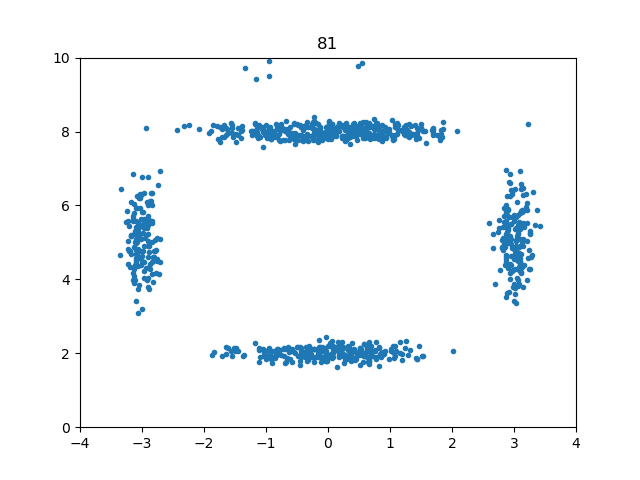}

\caption{We plot the samplers along updates. We have 100 updates in total. From left to right being the 1st, 21st, 41st, 61st, and 81st update. From up to down being the BDEC, BDLS, and ALPS. The range of plot is set to be $[-4, 4] \times [0, 10]$.}
    \label{fig:particles2}
\end{figure}

Moreover, as illustrated in Figure~\ref{fig:particles2}, BDEC exhibits a significantly faster exploration rate compared to BDLS. While BDEC and ALPS share the same exploration component, our scheme demonstrates superior efficiency in propagating information about newly discovered modes back to the original temperature.

To quantitatively compare the exploration process of different methods, we define a metric called the \textbf{exploration rate}:
\[
Z_t := \int_{\Omega} \pi(x) \mathds{1}_{\rho_t > 0}(x) \, dx,
\]
which represents the weighted sum of \( \pi \) over the support of \( \rho_t \). 
Denote the support of $\rho_t$ as $\Omega_t := \overline{\{x: \rho_t(x) > 0 \}}$, the exploration rate can also be written as 
\begin{equation*}
    Z_t = \int_{\Omega_t} \pi(x) \,dx. 
\end{equation*}
We plot \( Z_t \) as a function of the iteration count.

\begin{figure}[tbhp]
    \centering
    \subfloat[Evolution of $Z_t$]{\label{fig:2a}\includegraphics[width=0.4\textwidth]{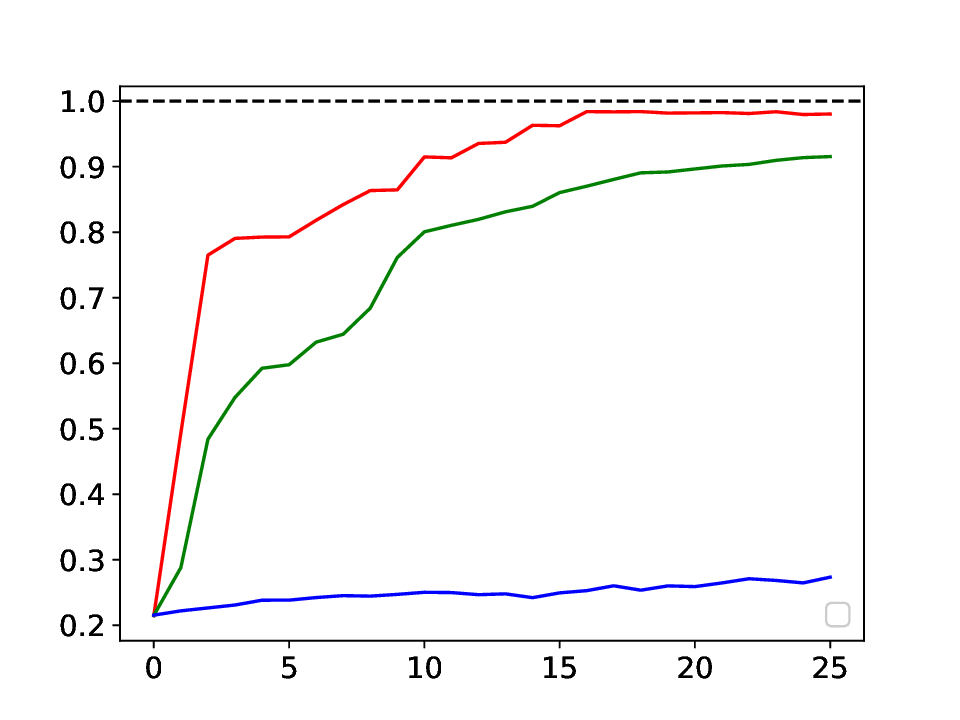}}
    \subfloat[Evolution of $\frac{1}{Z_t} - 1$]{\label{fig:2b}\includegraphics[width=0.4\textwidth]{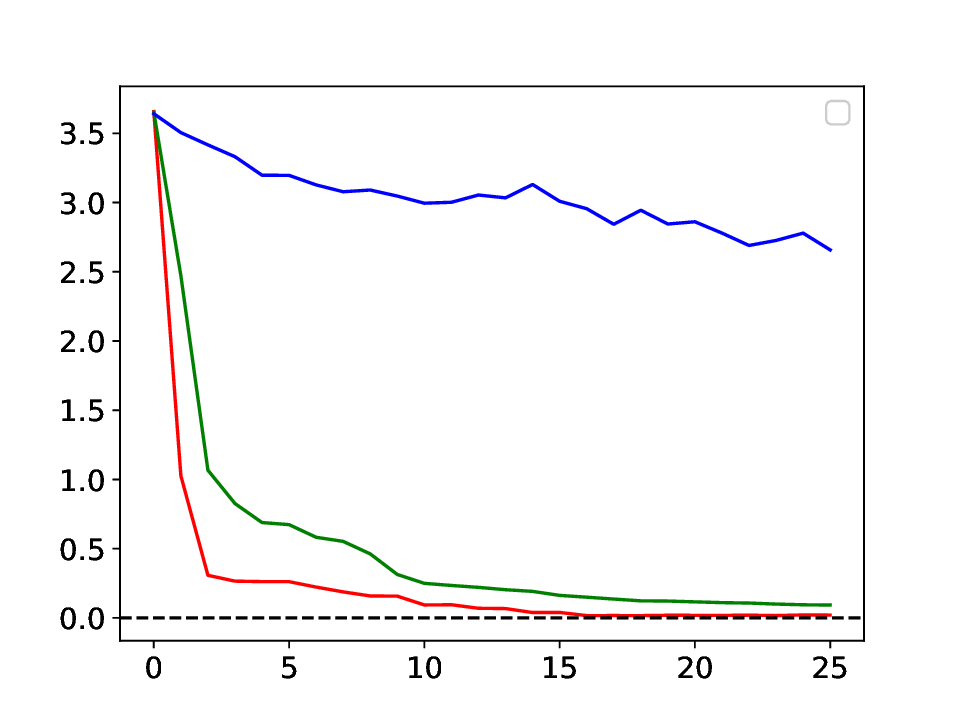}}
    \caption{Red, BDEC; Green, ALPS; Blue, BDLS.}
\label{figure:Z}
\end{figure}

To approximate \( Z_t \), we use the formula:
\begin{equation*}
    Z_t = \int_{\Omega_t} \pi(x) \, dx \approx \frac{1}{N} \sum_{n=1}^N \mathds{1}\{\omega_n \in \Omega_t\},
\end{equation*}
where \( \omega_n \overset{\mathrm{iid}}{\sim} \pi \). Since this example involves a mixture of four equally weighted Gaussian distributions, we have prior access to \( \omega_n \). Following Equation~\ref{itemrho}, we approximate the domain as \( \Omega_t \approx \bigcup_{i=1}^N \mathbb{B}(x_i^t, 4h) \), where \( \mathbb{B}(x_i^t, 4h) \) denotes the ball of radius \( 4h \) centered at \( x_i^t \). For our experiments, we set \( K = 20000 \), \( h = 0.05 \), and estimate \( Z_t \) at 25 intervals (see Figure~\ref{fig:2a}).

Also notice that, the \( \chi^2 \)-divergence $D_t$ is lower bounded by \( \frac{1}{Z_{t}} - 1 \): 
\begin{equation*}
    (D_t + 1) \cdot Z_t = \int \frac{\rho_t^2}{\pi} dx \cdot \int \pi \mathds{1}_{\rho_t(x) > 0} dx \geq \left( \int \rho_t \mathds{1}_{\rho_t(x) > 0} dx \right)^2 = 1
\end{equation*}
To complement the analysis, we also plot \( \frac{1}{Z_t} - 1 \) at each step (see Figure~\ref{fig:2b}).

BDEC and ALPS have faster convergence speed than BDLS. This should be credited to the construction of a tempered level and the exploration component. Also notice that $Z_t$ in BDEC process grows faster than in ALPS. This means, the information is efficiently passed to the target level by the MH step, so constructing annealed level is unnecessary. We will further discuss this point in another example \ref{subsection:Example3}.

\subsection{Example 2: 2D Seemingly-Unrelated Regression Model}

We also experiment on a 2-dimensional SUR model from \cite{tawn2021annealed} with parameters given in \cite{drton2004multimodality}. We omit the introduction of the background of the SUR model and give the target distribution directly:
\begin{equation}
\pi(\bm{x}) \propto e^{V(\bm{x})}, 
V(\bm{x}) = N \log(2\pi) + \frac{N}{2} \log(|\bm{\Sigma}|) + N
\end{equation}
after calculating the parameters given in \cite{drton2004multimodality}, we can obtain the explicit formula:
$$
\bm{\Sigma} =  \begin{pmatrix} 7.70{x_1}^2-19.27x_1 + 21.09 & 5.11x_1 x_2 - 3.42x_1 - 3.51x_2 + 23.52 \\ 5.11x_1 x_2 - 3.42x_1 - 3.51x_2 + 23.52 &27.31{x_2}^2 - 97.40x_2 + 114.19 \end{pmatrix}
$$
Our target distribution $\pi$ has 2 modes: $\mu_1 = (0.78,1.54)^T, \mu_2 = (2.76,2.50)^T$. We keep $1000$ samplers and they start from $\mu_0 = (1.25,1.78)^T$ with a standard Gaussian noise. \par
\begin{figure}[tbhp]
\centering
\subfloat[Estimating $\mathbb{E}(|x|)$]{\label{fig:301}\includegraphics[width=0.33\textwidth]{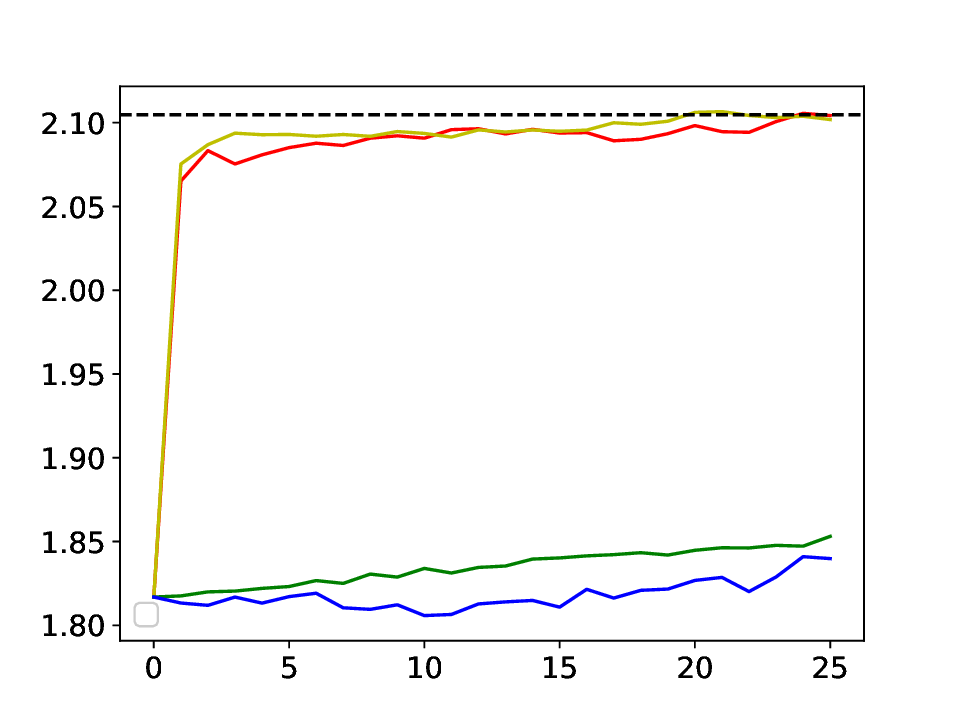}}
\subfloat[Estimating $\mathbb{E}(y)$]{\label{fig:302}\includegraphics[width=0.33\textwidth]{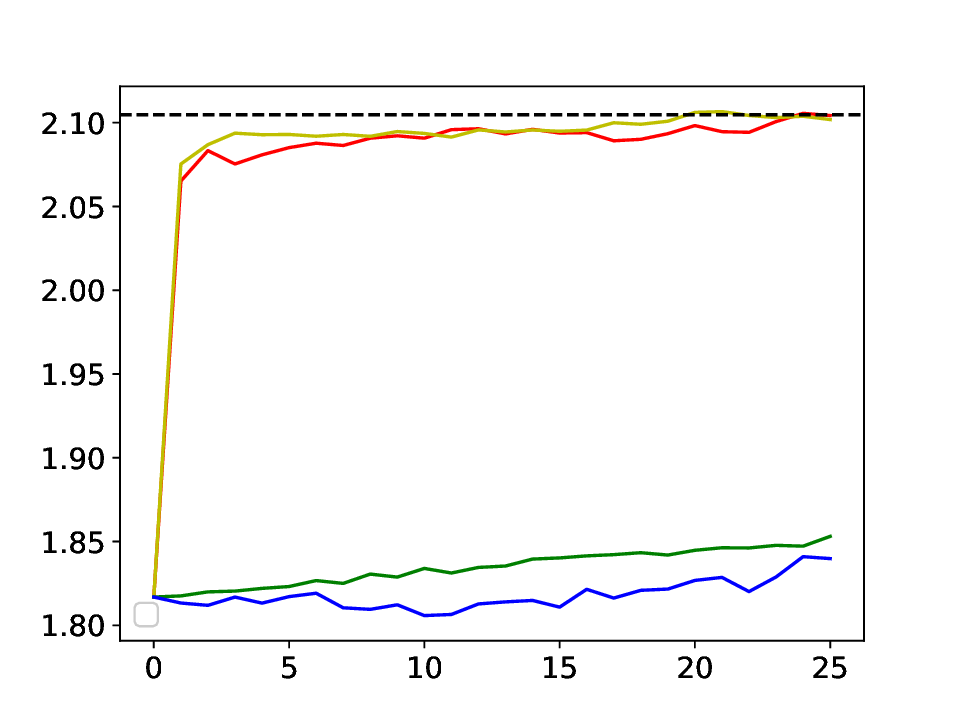}}
\subfloat[Estimating $\mathbb{E}(x^2/3+y^2/5)$]{\label{fig:303}\includegraphics[width=0.33\textwidth]{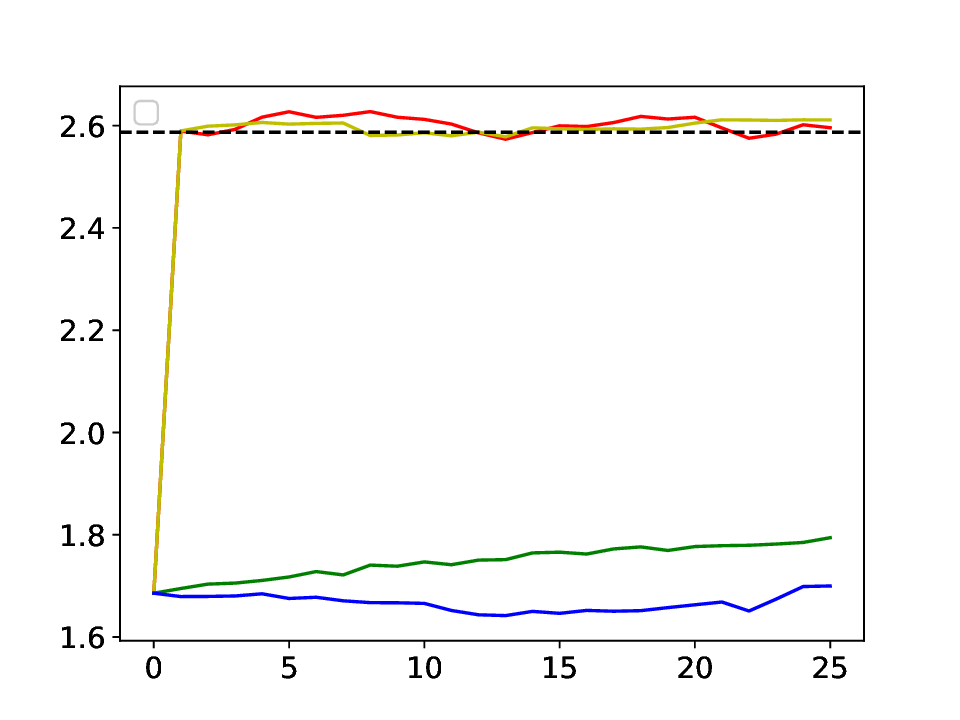}}
\caption{Errors of estimating $\mathbb{E}(f)$ for three different $f$ in Example 2. Red, estimation of BDEC; Yellow, LEC; Green, ALPS; Blue, the true expectation value. We plot the estimations for 30 total iterations with each iteration including 5 updates.}
\label{fig:example3}
\end{figure}

 
As the range scale is smaller this time, we set time step to be $\Delta t = 0.0005$ and kernel width $h = 0.05$. Still, we choose number of total iterations $J = 25$ and number of within-temperature moves $T = 4$ for BDEC, which equals $100$ total iterations and $3$ within-temperature moves for ALPS (because the swap should be counted as one update). We choose hotter temperature $\beta_{hot} = 0.05$ for both algorithms and a batch size $B = 12$ for the Exploration Component. We run ALPS using the same annealing scheme $(1,\sqrt{10},10)$ from \cite{tawn2021annealed}. Still, we compare the absolute error of estimating $\mathbb{E}(f)$ for different $f$. As depicted in Figure~\ref{fig:example3}, the outcomes of the BDEC and LEC algorithms display remarkable similarity, with both converging swiftly towards the target value. In contrast, the ALPS and BDLS algorithms demonstrate slower performance. This observation aligns with the fact from \cite{tawn2021annealed} that the ALPS algorithm has a notably long burn-in phase.

Additionally, we visualize the exploration processes of these algorithms by plotting the samplers across various iterations. The similarity between the samplers in the first and second rows reinforces the conclusion that, under this 2D two-mode setting, the birth-death process offers limited advantage. Conversely, the samplers in the third and fourth rows looks like plain and slow diffusion. A comparative analysis between the first and third rows reveals significant differences introduced by the addition of an exploration component to the original BDLS algorithm, particularly in the early stages. Comparing the first and fourth rows highlights higher efficiency of BDEC against ALPS in absorbing information from the exploration component. The final line illustrates the samplers at the annealed level within the ALPS algorithm, which confirms that the two mode information was successfully identified and delivered to this annealed level. Consequently, the inefficiency of ALPS should be attributed to the fact that the swapping moves fail to transmit mode location information back to the original temperature level.

\begin{figure}[h!] 
\centering

\includegraphics[width=0.19\textwidth]{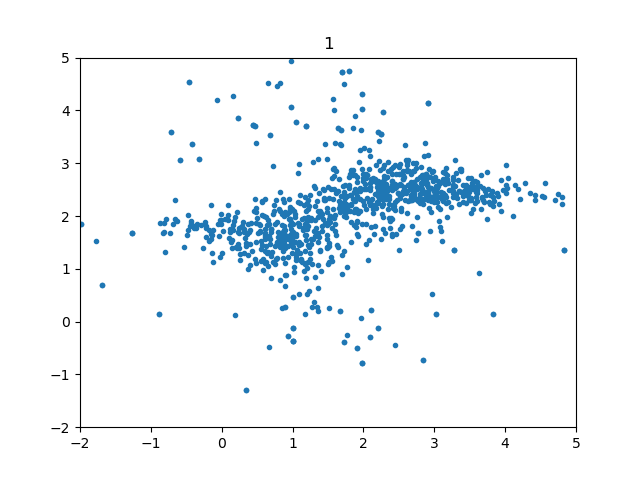}
\includegraphics[width=0.19\textwidth]{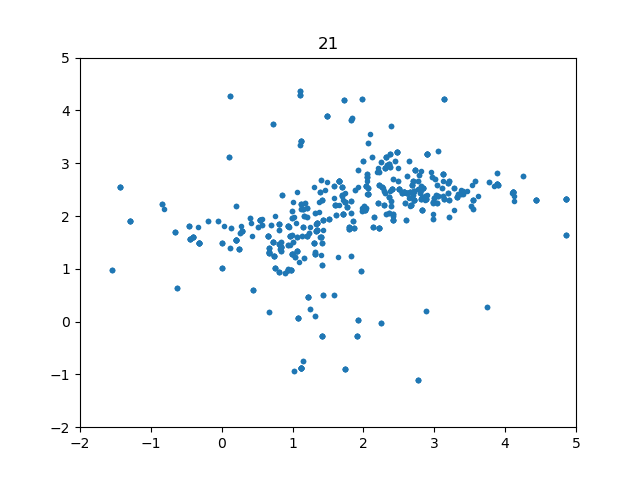}
\includegraphics[width=0.19\textwidth]{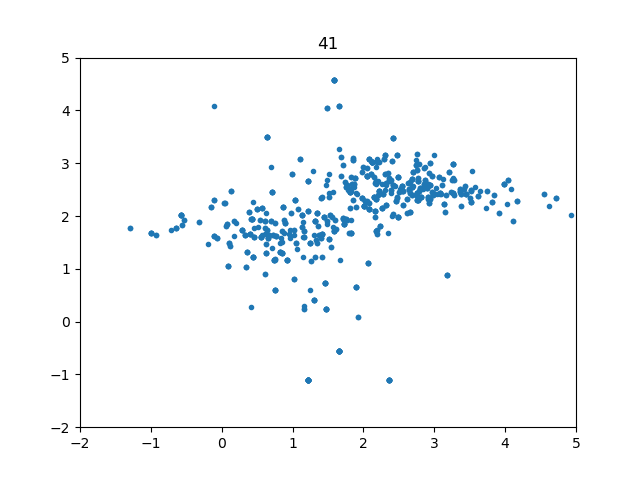}
\includegraphics[width=0.19\textwidth]{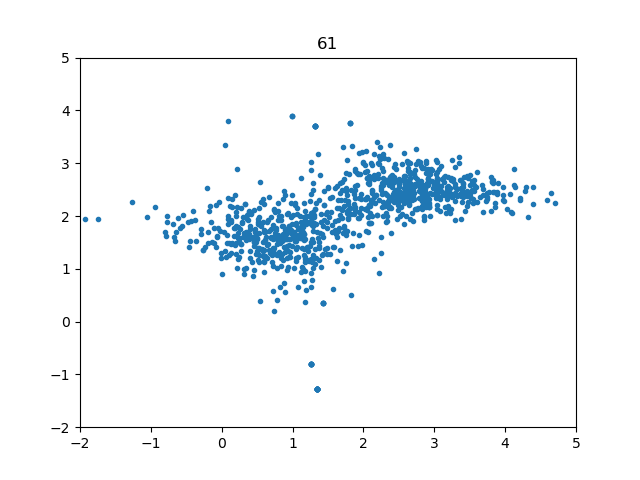}
\includegraphics[width=0.19\textwidth]{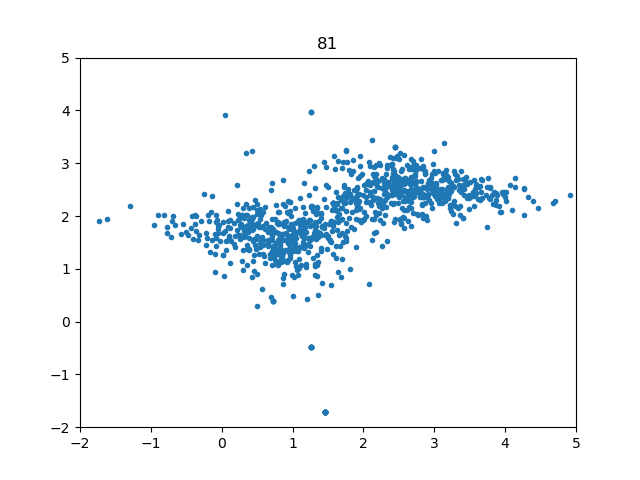}

\includegraphics[width=0.19\textwidth]{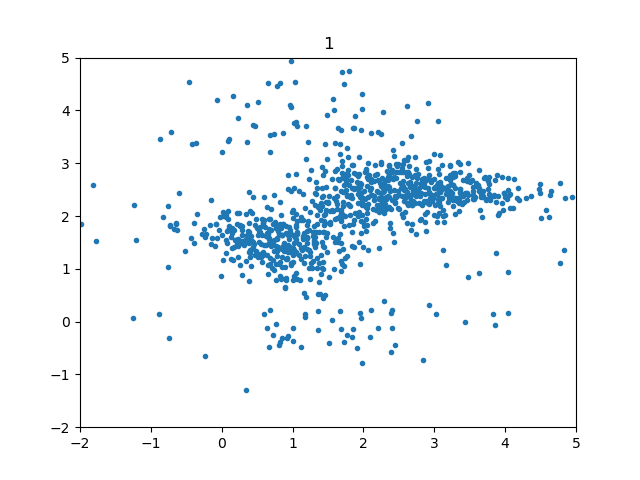} \includegraphics[width=0.19\textwidth]{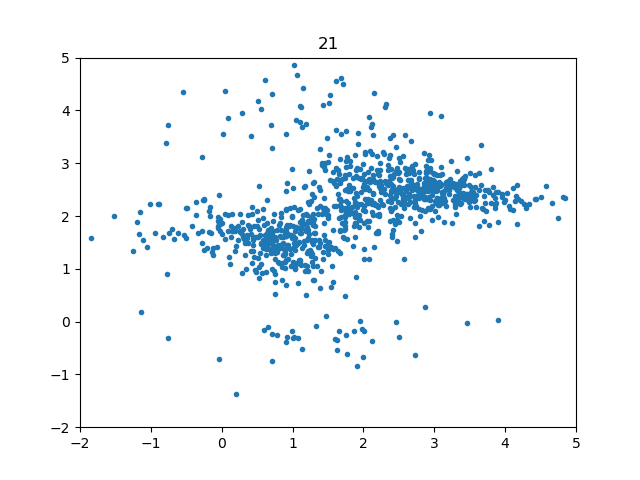}
\includegraphics[width=0.19\textwidth]{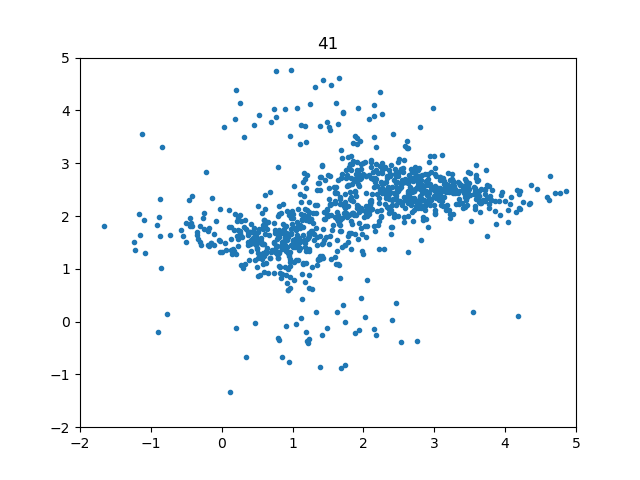}
\includegraphics[width=0.19\textwidth]{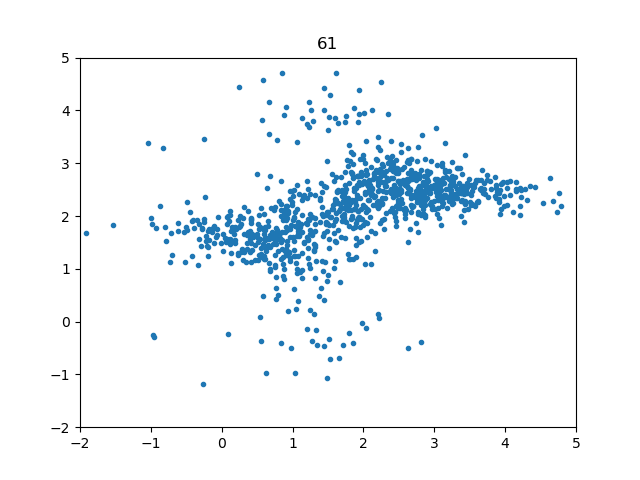}
\includegraphics[width=0.19\textwidth]{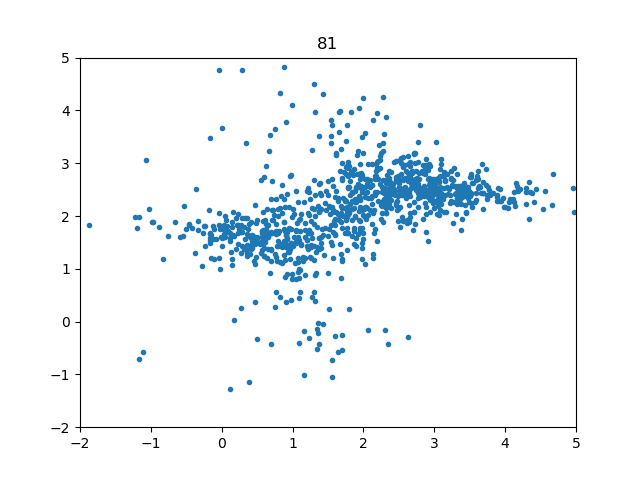}

\includegraphics[width=0.19\textwidth]{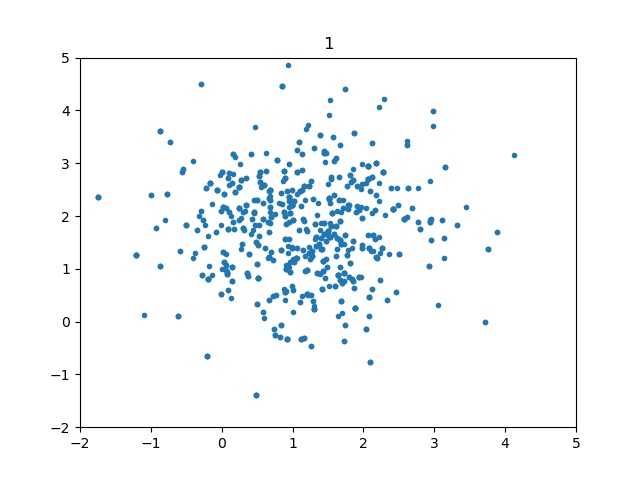}
\includegraphics[width=0.19\textwidth]{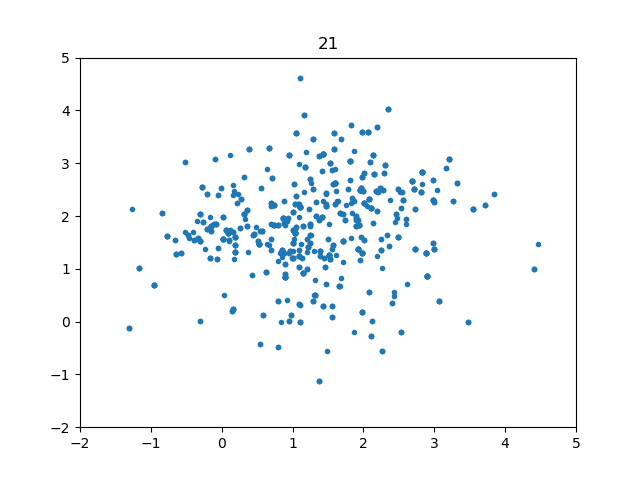}
\includegraphics[width=0.19\textwidth]{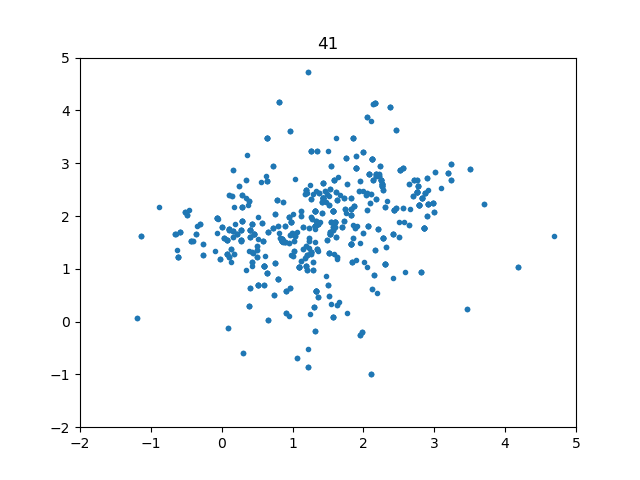}
\includegraphics[width=0.19\textwidth]{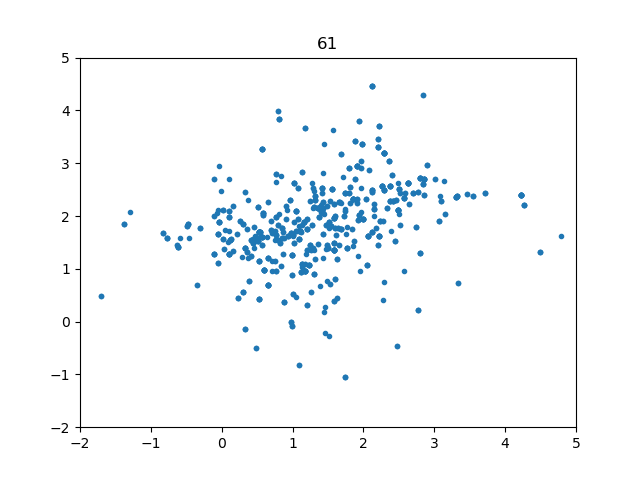}
\includegraphics[width=0.19\textwidth]{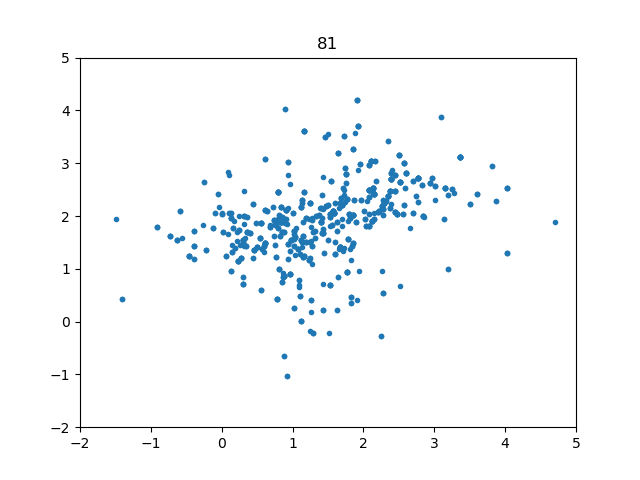}

\includegraphics[width=0.19\textwidth]{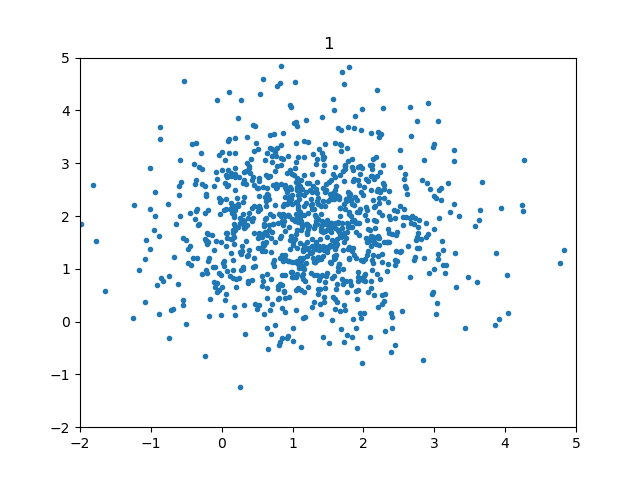}
\includegraphics[width=0.19\textwidth]{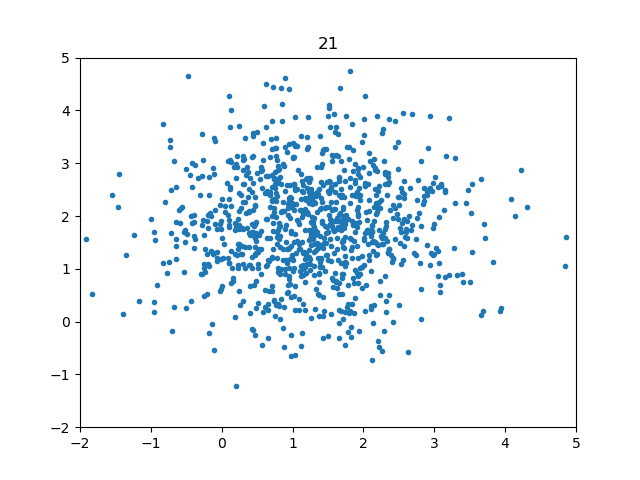}
\includegraphics[width=0.19\textwidth]{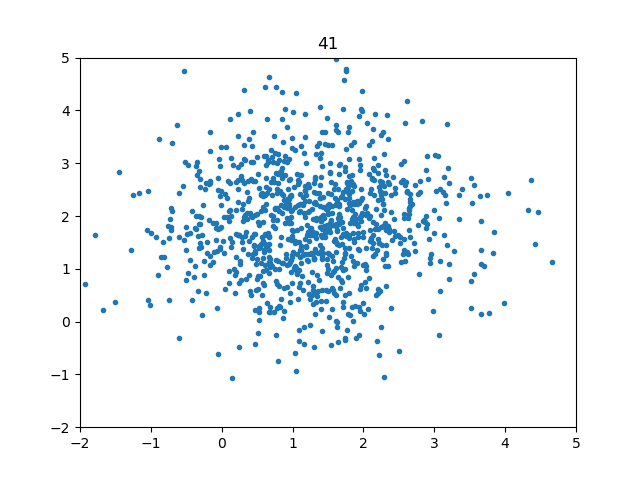}
\includegraphics[width=0.19\textwidth]{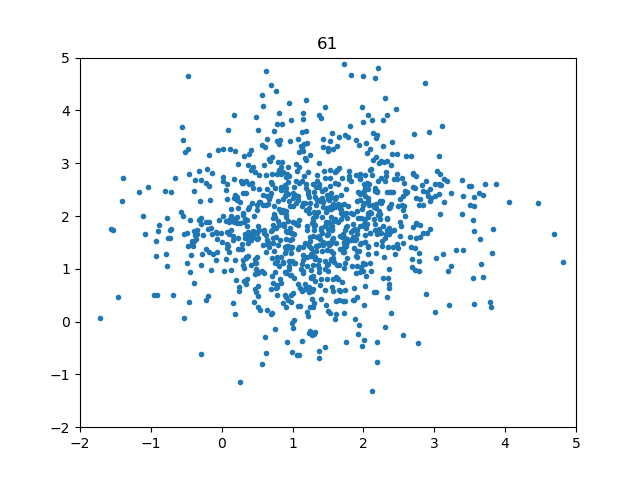}
\includegraphics[width=0.19\textwidth]{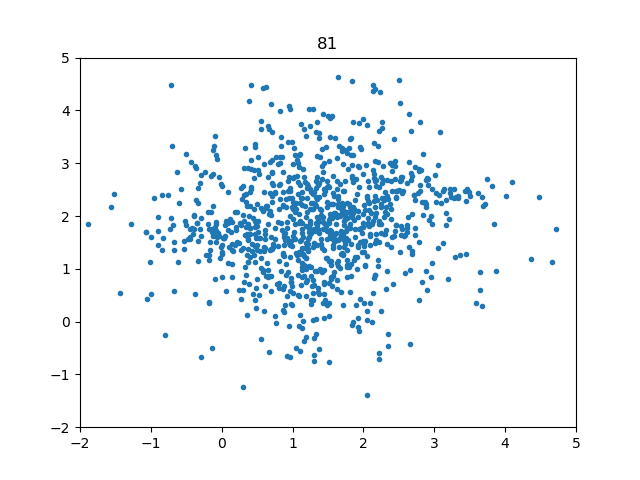}

\includegraphics[width=0.19\textwidth]{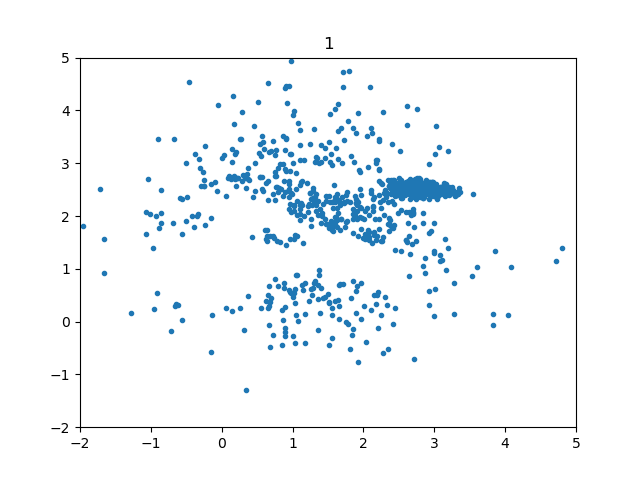}
\includegraphics[width=0.19\textwidth]{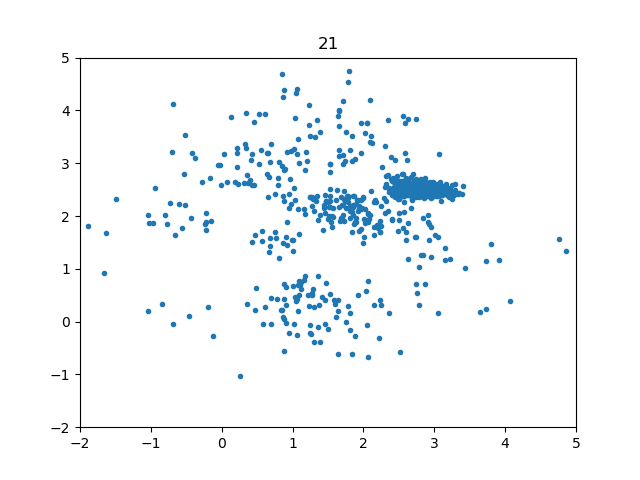}
\includegraphics[width=0.19\textwidth]{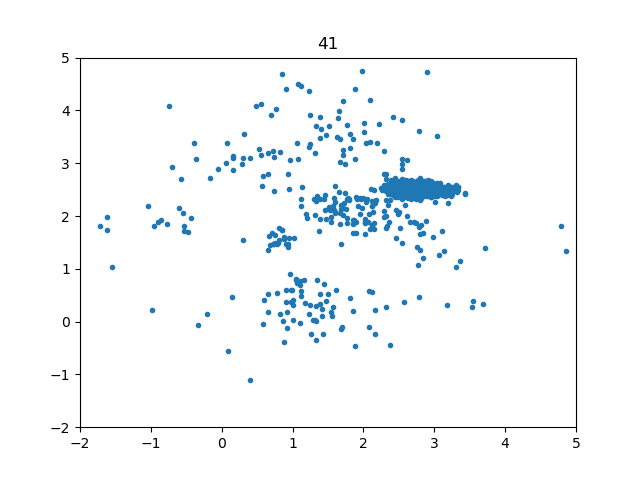}
\includegraphics[width=0.19\textwidth]{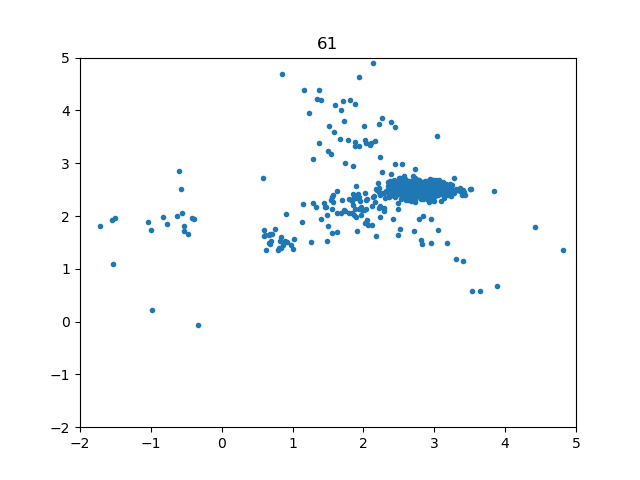}
\includegraphics[width=0.19\textwidth]{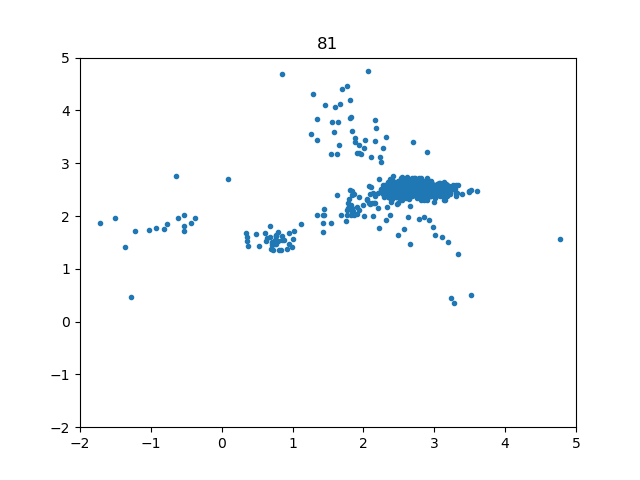}

\caption{We plot the samplers along updates. We have 100 updates in total. From left to right being the 1st, 21st, 41st, 61st, and 81st update. From up to down being the BDEC, LEC, BDLS, ALPS and the annealed level of ALPS. The range of the plot is set to be $[-2, 5] \times [-2,5]$.}
    \label{fig:particles3}
\end{figure}


\subsection{Example 3: 20D Problem}
\label{subsection:Example3}
For the last experiment, we run the algorithm in higher dimensions. The target distribution is a mixture of four well-separated 20-dimensional distributions with heterogeneous scaling used in \cite{tawn2021annealed}. The target distribution is given by 
\begin{equation}
\pi(x) \propto \sum_{k=1}^4 \prod_{j=1}^{20} \frac{2}{w_k}\phi(\frac{x_j-(\mu_k)_j}{w_k})\Phi(\alpha \frac{x_j-(\mu_k)_j}{w_k}) 
\end{equation}
where $\alpha = 10, \mu_1 = (20,20,...,20) = -\mu_2, \mu_3 = (-10,-10,...,-10,10,10,...,10) = -\mu_4, w_1 = w_2 = 1, w_3 = w_4 = 2$, and $\phi(\cdot),\Phi(\cdot)$ denote the density and CDF of standard Gaussian distribution. \par
We keep $1000$ samplers and they are initialized from the first mode $\mu_1$ with a standard Gaussian noise. We set time step $\Delta t = 0.005$ and a kernel bandwidth $h = 0.2$ in the birth-death process. We set the number of within-temperature moves $T = 4$ in each iteration for BDEC, which equals $3$ within-temperature moves each iteration for ALPS. To make a fair comparison with \cite{tawn2021annealed}, we kept the same parameters, with the warmest temperature being $\beta_{hot} = 0.00005$ for both algorithms and the annealing scheme being ${1,4,16,64,256}$. \par
\begin{figure}[tbhp]
\centering
\includegraphics[width=0.46\textwidth]{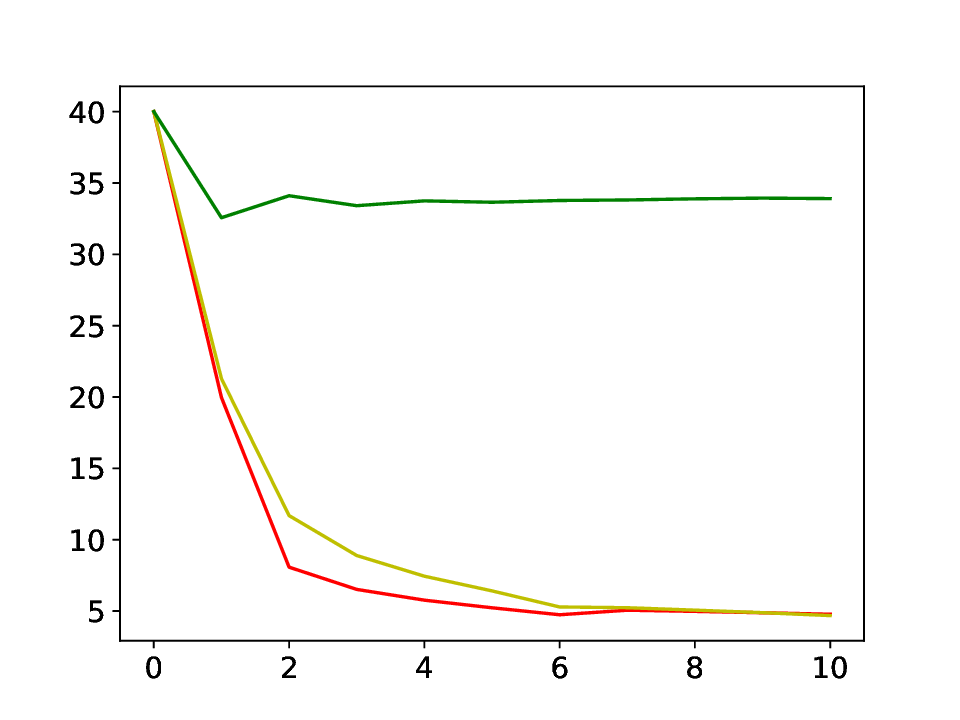}     
\caption{Absolute value of $\mathbb{E}(x_1 + x_2)$. Red, estimation of BDEC; Yellow, LEC; Green, ALPS. We plot the divergence for 10 total iterations with each iteration including 3 updates.}
\label{fig:example4_0}
\end{figure}

Notice that the target distribution $\pi$ is symmetric with respect to $x_j = 0$ for each axis $x_j$. As a result, the expectation of $x_j$ for each $j = 1, \ldots, 20$ should be exactly $0$. We evaluate the expectation of the first and second axes at each iteration and plot the results in Figure~\ref{fig:example4_0}.

To make the comparison clearer, we calculate KL-divergence of the first coordinate $\KL(\rho^1_t|\pi^1)$ inside each iteration, where $\rho^1_t,\pi^1$ denotes the marginal distribution of $\rho_t,\pi$ and $\rho_t = \sum_{i=1}^N \mathcal{N}(\cdot; x_i^t, h\bm{I})$ is the kernel distribution of $N$ samplers. Because the marginal distributions of all coordinates are heterogeneous, we believe the KL-divergence $\KL(\rho^i_t|\pi^1)$ can be representative of all $\KL(\rho^1_t| \pi^i), i = 1,2,...,20$. \par
The comparative analysis of the red and green lines clearly demonstrates that our BDEC scheme outperforms ALPS in terms of efficiency per iteration. To ensure a fair comparison, we examined the time required for a single iteration in both BDEC and ALPS. The CPU time for one iteration in BDEC is approximately 936.79 units, with the birth-death process accounting for 6.81\%. In contrast, the CPU time for an iteration in ALPS is around 1290.29 units, which is 37.74\% more than that of our BDEC algorithm. This comparison conclusively shows that BDEC is more time-efficient.
Furthermore, in the BDEC scheme, the birth-death process is applied after the initial 10 iterations. The comparison between the red and yellow lines illustrates that the inclusion of the birth-death process enhances the convergence rate, leading to faster achievement of the desired results. \par

\begin{figure}[tbhp]
\centering
\includegraphics[width=0.47\textwidth]{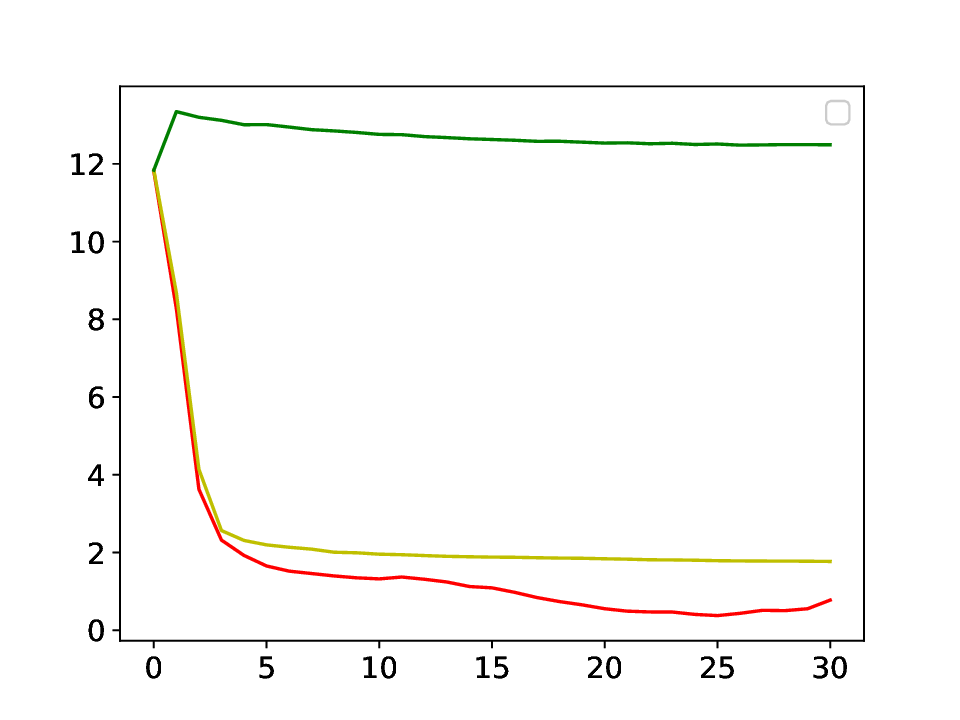}     
\caption{Convergence of KL-divergence of the $1^{st}$ coordinate. Red, estimation of BDEC; Yellow, LEC; Green, ALPS. We plot the divergence for 30 total iterations with each iteration including 3 updates.}
\label{fig:example4}
\end{figure}

The experimental results suggest that the annealing component in the ALPS algorithm may not be as useful as anticipated. In ALPS, annealing is designed to mitigate skewness issues: it achieves this by better approximating a skewed target distribution with a mixture of Gaussian distributions at lower temperatures. This enhancement increases the acceptance rate of the random walk Metropolis-Hastings algorithm, facilitating the transfer of new mode information to the Markov chains. However, its approach requires maintaining multiple levels of Markov chains simultaneously, leading to significant time consumption.
In contrast, the BDEC scheme approaches this challenge differently. Although the acceptance rate in BDEC at original temperature might be lower due to skewness, once a sampler approaches a new mode's vicinity, the birth-death process efficiently transports multiple samplers to this new region almost instantaneously. This strategy effectively addresses skewness issues without the added computational burden associated with multiple Markov chains. Overall, BDEC proves to be an efficient method for dealing with both multimodal and skewed distributions.

\section{Limitations and Further Work} 
\label{sec:conclusions}

Our results demonstrate that BDEC has outstanding performance, combining the advantages of both BDLS and ALPS. However, there are limitations and space for further discussion:
\begin{itemize}
    \item Applications on practical settings: the experiment settings we use are artificial and have few numbers of modes. Distributions in real-world problems may have higher dimensions, larger scale, and more modes. 
    \item Computational cost: second-order derivatives were applied to decide whether a mode is new \eqref{eq:distance}, and also applied in Gaussian approximation \eqref{eq:defproposal}. Besides, Kernel density estimation (KDE) is used during the birth-death process, but KDE is afflicted by the curse of dimensionality. This is a common problem across interacting particle systems, including SVGD \cite{liu2016stein} and Ensemble MCMC \cite{lindsey2022ensemble}.
    \item Notice that the exponential convergence rate in Theorem~\ref{Theorem_main} is $2(1 - \delta) + 2c\tau_2$. This highlights the effort of BD and EC, but ignores the contribution of Langevin sampling. There might be sharper analysis that takes ULA into consideration. Also notice that the convergence speed can be astoundingly fast if $\tau_1, \tau_2 \to \infty$. However, in practice, there is no specific algorithm corresponds to this limiting situation. There is hope for deriving new sampling methods.
\end{itemize} \par
If we omit the technical details, the idea we present is simply \textit{look before you leap}. `Look' refers to letting $Y$ explore the unknown space and collect useful information as much as possible. `Leap' refers to $X$ sampling the target distribution after having abundant information. Under this framework, we are actually proposing a toolbox for sampling multimodal distributions. It is free to substitute the Langevin Dynamics by other well-known Markov Chain Monte Carlo approaches. We can replace the Metropolis-Hastings step for the first few steps by directly inserting new modes for the sake of simplicity. Also, we can apply the exploration component for multiple times instead of just one before $X$ starts to move. In all, we are looking for variations of BDEC and their applications on wider range of problems.

\newpage

\appendix

\section{Proof of Lemmas}
\label{ap_sec:proof_lem}
\subsection{Proof of Lemma~\ref{lem:G_t_lower_1}}
\label{ap_sec:G_t_lower_1}

\begin{proof}
Let us recall the dynamic of $g_t$, see \eqref{dynamicofg}, 
    \begin{equation*}
        \partial_t g_t  = -\nabla V \cdot \nabla g_t + \Delta g_t -  \tau_1(g_t - \bar{g}_t)g_t + \tau_2 \int_{\Omega} R(x,y)(g_t(y) - g_t(x))\pi(y)dy 
    \end{equation*}
    Define the function 
    \begin{equation*}
        h^{\epsilon}_t(x) = (g_t - 1) e^{(c\tau_2 - \epsilon) t}
    \end{equation*}
    where $\epsilon$ is a positive parameter. 
    \begin{align*}
        \partial_t h_t^{\epsilon}  = & \partial_t \left((g_t- 1)e^{(c\tau_2 - \epsilon) t}\right) \\
         = & [\partial_t g_t  + (c\tau_2 - \epsilon)(g_t - 1)]  e^{(c\tau_2 - \epsilon) t} \\
        = & [-\nabla V \cdot \nabla g_t + \Delta g_t - \tau_1 (g_t - \bar{g}_t)g_t \\
        & + \tau_2 \int_{\Omega} R(x,y)(g_t(y) - g_t(x))\pi(y)dy   + (c\tau_2 - \epsilon)(g_t - 1)]e^{(c\tau_2 - \epsilon) t}
    \end{align*}
    We want to prove that the minimum of $h^\epsilon_t$ over $\Omega \times [0, t_0]$ can only be reached at $t = 0$. 
    Suppose $h^\epsilon_t$ reaches minimum at $(x_0, t_0)$ for $t_0 > 0$. We give a lower bound for the right hand side at $x = x_0, t = t_0$. \\
    First, 
    $g_{t_0}(x)$ has a local minimum at $x = x_0$. According to periodic boundary condition (Assumption~\ref{asp:periodic}),  we have $\nabla g_{t_0}(x_0) = 0$, $\Delta g_{t_0}(x_0) \geq 0$. 
    Second, as $g_{t_0}(x_0)$ is smaller than the average of $g_{t_0}(\cdot)$, which means exactly $g_t \leq \bar{g}_t$, the second term $- (g_t - \bar{g}_t) g_t$ is also non-negative. 
    Finally, assumption~\ref{asp:explored} implies $R(x,y) = \min\{\hat{g}_t(x), \hat{g}_t(y) \} \geq c$. Together with $g_t(y) \geq g_t(x_0)$ at $t = t_0$, we get 
    \begin{equation*}
        \tau_2 \int_{\Omega} R(x,y)(g_t(y) - g_t(x))\pi(y)dy \geq c \tau_2 \int_{\Omega} (g_{t_0}(y) - g_{t_0}(x_0))\pi(y)dy = c \tau_2 (1 - g_{t_0}(x_0))
    \end{equation*}
    Therefore the bracket in the right hand side is no smaller than $c \tau_2 (1 - g_{t_0}(x_0)) + (c \tau_2 - \epsilon) (g_{t_0}(x_0) - 1)  = \epsilon (1 - g_{t_0}(x_0))$. However, 
    \begin{equation*}
        0 \geq \partial_t h^{\epsilon}_t(x_0) |_{t = t_0} \geq \epsilon (1 - g_{t_0}(x_0)) e^{(c\tau_2 - \epsilon) t_0} > 0, \text{ contradiction! }
    \end{equation*}
    Therefore, $h^\epsilon_t(x)$ can only reaches its minimum at $t = 0$, which implies 
    \begin{equation*}
        (g_t(x) - 1) e^{(c\tau_2 - \epsilon) t} \geq \inf_{x \in \Omega} g_0(x) - 1 \geq -1 \Longrightarrow g_t(x) \geq 1 - e^{ - (c\tau_2 - \epsilon) t}
    \end{equation*}
    Since it holds for any positive $\epsilon > 0$, we conclude with \eqref{eq:inf_g}. 
\end{proof}

\subsection{Proof of Lemma~\ref{lem:G_t_lower_2}}
\label{ap_sec:G_t_lower_2}

\begin{proof}
We proceed the proof with $\tau_1 = 1$. From the definition of $\bar{g}_t$ we have 
$\bar{g}_t = \int \rho_t^2/\pi \cdot \int \pi \geq \left( \int \rho_t \right)^2 = 1$. Using this, the dynamics of $g_t$ satisfy
    \begin{equation} 
    \label{eq:dyn_g_lower}
    \begin{aligned}
\partial_t g_t & = -\nabla V \cdot \nabla g_t + \Delta g_t -  \tau_1(g_t - \bar{g}_t)g_t + \tau_2 \int_{\Omega} R(x,y)(g_t(y) - g_t(x))\pi(y)dy \\
& \geq -\nabla V \cdot \nabla g_t + \Delta g_t - \tau_1(g_t - 1)g_t + \tau_2 \int_{\Omega} R(x,y)(g_t(y) - g_t(x))\pi(y)dy.
    \end{aligned}
\end{equation}
We first claim that $g_t(x)$ can only reach its minimum at $t = 0$. Define $g^\epsilon_t(x) = g_t(x) + \epsilon t$ where $\epsilon > 0$. Then, $\nabla g^\epsilon_t = \nabla g_t, \Delta g^\epsilon_t = \Delta g_t$, and $\partial_t g^\epsilon_t = \partial_t g_t + \epsilon$. Substituting \eqref{eq:dyn_g_lower} into the expression of $\partial_t g^\epsilon_t$, we get 
\begin{equation*}
    \partial_t g^\epsilon_t \geq \epsilon - \nabla V \cdot \nabla g^\epsilon_t + \Delta g^\epsilon_t + \tau_1(1 - g_t) g_t + \tau_2 \int_{\Omega} R(x,y)(g^\epsilon_t(y) - g^\epsilon_t(x))\pi(y)dy.
\end{equation*}
Assume, for contradiction, 
the minimum of $g^\epsilon_t$ over $\Omega \times [0,t_0]$ is reached at $x_0$ and $t_0 > 0$. Then,  the left hand side $\partial_t g^\epsilon_t(x_0) \leq 0$. \\
For the right hand side, we still have $\nabla g^\epsilon_t(x_0) = 0, \Delta g^\epsilon_t(x_0) \geq 0$ according to the periodic boundary condition. Second, notice that $g^\epsilon_{t_0}(x_0) \leq g^\epsilon_{t_0}(x)$ implies $g_{t_0}(x_0) \leq g_{t_0}(x)$. Thus, $g_{t_0} (x_0) \leq 1$, and the term $- ( g_{t_0}(x_0) - 1) g_{t_0} (x_0) \geq 0$. 
Moreover, $g^\epsilon_{t_0}(y) \geq g^\epsilon_{t_0}(x_0)$ for all $y$ and $R(x,y) \geq 0$, so the third integral term is also non-negative. 
Therefore:
\begin{equation*}
\partial_t g_{t_0}^{\epsilon}(x_0)
\geq \epsilon > 0,
\end{equation*}
which contradicts $\partial_t g_{t_0}^{\epsilon}(x_0) \leq 0$. 
Therefore, $g_t(x) + \epsilon t \geq \inf_{x \in \Omega} g_0(x)$. Since $\epsilon > 0$ can be arbitrarily small, we conclude $g^t(x) \geq e^{-M}$.  

Now let us define a new function 
\begin{equation*}
    \eta_{t} = \log\frac{g_t}{1 - g_t},
\end{equation*}
with $\eta_t = + \infty$ if $g_t \geq 1$. We then claim that $\eta_t - t$ can only reaches minimum at $t = 0$. \\
Define $\eta^{\epsilon}_t = \eta_t - t + \epsilon t$ and suppose, for contradiction,  
$\eta^{\epsilon}_t$ reaches its minimum at $x_0$ and $t_0 > 0$. Then, $g_{t_0}(x_0)$ is strictly smaller than $1$. According to continuity, there exists a region around $(x_0, t_0)$ such that $g_t(x) \in (0,1)$. 
We rewrite $\eqref{eq:dyn_g_lower}$ as
\begin{equation}
\label{eq:dyn_eta_lower}
    \frac{\partial_t g_t}{(1- g_t)g_t} - 1 \geq - \nabla V \cdot \frac{\nabla g_t}{(1- g_t)g_t} + \frac{\Delta g_t}{(1- g_t)g_t}  + \frac{\tau_2 \int_{\Omega} R(x,y)(g_t(y) - g_t(x))\pi(y)dy}{(1- g_t)g_t} 
\end{equation}
and consider the right hand side at $t = t_0$ and $x = x_0$.
Since $\log (a/ 1-a)$ is a monotone increasing function of $a$ on $(0,1)$, $g_{t_0}(x)$ also has a local minimum at $x_0$, which further implies $\nabla g_{t_0}(x_0) = 0$, $\Delta g_{t_0}(x_0) \geq 0$. As previously discussed, the integral term is non-negative. So we have 
\begin{equation*}
    \partial_t \eta^\epsilon_{t_0}(x_0) - \epsilon = \frac{\partial_t g_{t_0}(x_0)}{(1- g_{t_0}(x_0))g_{t_0}(x_0)} - 1 \geq 0, 
\end{equation*}
which contradicts $\partial_t \eta^\epsilon_{t_0}(x_0) \leq 0$.
Accordingly, we have $\eta_t(x) - t + \epsilon t \geq \inf_{x \in \Omega}\eta_0(x)$ for all $\epsilon > 0$. So $\eta_t(x) - t \geq \inf_{x \in \Omega}\eta_0(x)$. From the definition of $\eta_t$:
\begin{align*}
    & \log \frac{g_t(x)}{1 - g_t(x)} \geq e^{t} \log \frac{e^{-M}}{1 - e^{-M}} \\
    \Longrightarrow & g_t(x) \geq \frac{1}{e^{-t}(e^M - 1) + 1}. 
\end{align*}
This completest the proof. 
\end{proof}

\newpage

\bibliographystyle{abbrv}
\bibliography{ref_siam.bib}

\end{document}